\documentclass{nature_hack}

\usepackage{deluxetable}
\usepackage{graphicx}	
\usepackage{amsmath}	
\usepackage{amssymb}
\usepackage{times}
\usepackage{latexsym}
\usepackage{fnpos}
\usepackage{scrextend}
\usepackage{scalerel}
\usepackage{hyperref}
\usepackage{tabularx}
\usepackage{xspace}
\usepackage{enumerate}
\usepackage{longtable}
\usepackage{lscape}

\usepackage[a4paper, total={6in, 8in}]{geometry}
 \newcommand*\aap{Astron. Astrophys.}

\newcommand*\actaa{Acta Astron.}
\newcommand*\aj{Astron. J.}

\newcommand*\apj{Astrophys. J.}
\newcommand*\apjl{Astrophys. J. l.}

\newcommand*\apjs{Astrophys. J. S.}

\newcommand*\apss{Astrophys.~Sapce~Science}
\newcommand*\araa{An. Rev. Astron. Astrophys.}

\newcommand*\jrasc{JRASC}

\newcommand*\mnras{Mon. Not. R. Astron. Soc.}

\newcommand*\nat{Nature}

\newcommand*\pasp{Publications of the Astronomical Society of the Pacific}

\newcommand*\prl{Phys.~Rev.~Lett.}
\newcommand*\procspie{Proc.~SPIE}

\newcommand*\rmxaa{Rev. Mexicana Astron. Astrofis.}

\newcommand*\ssr{Space~Sci.~Rev.}
\newcommand*\zap{ZAp}
\date{}

\newcommand{\eal}[2]{\ifmmode{\mathrm{#1\,#2}}\else{#1\textsc{$\,$\lowercase{#2}}}\fi\xspace}
\newcommand{\feal}[2]{\ifmmode{\mathrm{#1\,#2}}\else{[#1\textsc{$\,$\lowercase{#2}}]}\fi\xspace}
\newcommand{\hfeal}[2]{\ifmmode{\mathrm{#1\,#2}}\else{#1\textsc{$\,$\lowercase{#2}}]}\fi\xspace}

\newcommand{\beginmain}{%
        \setcounter{figure}{0}
        \renewcommand{\figurename}{Figure}
        \renewcommand{\thefigure}{\arabic{figure}}%
     }

\newcommand{\beginsupplement}{%
        \setcounter{table}{0}
        \renewcommand{\tablename}{Supplementary\,\,Table}
        \renewcommand{\thetable}{\arabic{table}}%
        \setcounter{figure}{0}
        \renewcommand{\figurename}{Supplementary\,\,Figure}
        \renewcommand{\thefigure}{\arabic{figure}}%
     }

\title{Direct evidence for shock-powered optical emission in a nova}

\author{Elias Aydi$^{1*}$, Kirill V.\ Sokolovsky$^{1,2,3*}$, Laura Chomiuk$^{1*}$, Elad Steinberg$^{4,5}$, Kwan Lok Li$^{6,7}$, Indrek  Vurm$^{8}$, Brian D.\ Metzger$^{4}$, Jay Strader$^{1}$, Koji Mukai$^{9,10}$, Ond\v{r}ej Pejcha$^{11}$, Ken J.\ Shen$^{12}$, 
Gregg A.\ Wade$^{13}$, Rainer Kuschnig$^{14}$, Anthony F.~J.\ Moffat$^{15}$, Herbert Pablo$^{16}$, Andrzej Pigulski$^{17}$, Adam Popowicz$^{18}$, Werner Weiss$^{19}$, Konstanze Zwintz$^{20}$, 
Luca Izzo$^{21}$, Karen R.\ Pollard$^{22}$, Gerald Handler$^{23}$, Stuart D.\ Ryder$^{24}$, Miroslav D.\ Filipovi\'c$^{25}$, Rami Z. E.\ Alsaberi$^{25}$, Perica Manojlovi\'c$^{25}$, Raimundo~Lopes de Oliveira$^{26,27}$, Frederick M.\ Walter$^{28}$, Patrick J.\ Vallely$^{29}$, David A.~H.\ Buckley$^{30}$, 
Michael J.~I.\ Brown,$^{31}$,  Eamonn J.\ Harvey$^{32}$, Adam Kawash$^{1}$, Alexei Kniazev$^{30,33,34}$, Christopher S.\ Kochanek$^{29}$, Justin Linford$^{35,36,37}$, Joanna Mikolajewska$^{23}$, Paolo Molaro$^{38}$, Marina Orio$^{39,40}$, Kim L.\ Page$^{41}$, Benjamin J.\ Shappee$^{42}$ and Jennifer L.\ Sokoloski$^{4}$}

\begin{document} 
\spacing{1}
\maketitle
\begin{affiliations}
\item Center for Data Intensive and Time Domain Astronomy, Department of Physics and Astronomy, Michigan State University, East Lansing, MI 48824, USA

\item Sternberg Astronomical Institute, Moscow State University, Universitetskii pr. 13, 119992 Moscow, Russia 

\item Astro Space Center of Lebedev Physical Institute, Profsoyuznaya St. 84/32, 117997 Moscow, Russia

\item Columbia Astrophysics Laboratory and Department of Physics, Columbia University, New York, NY 10027, US

\item Racah Institute of Physics, Hebrew University, Jerusalem 91904, Israel

\item Department of Physics, UNIST, Ulsan 44919, Korea

\item Institute of Astronomy, National Tsing Hua University, Hsinchu 30013, Taiwan

\item Tartu Observatory, University of Tartu, T\~{o}ravere 61602, Tartumaa, Estonia

\item CRESST and X-ray Astrophysics Laboratory, NASA/GSFC, Greenbelt, MD 20771, USA

\item Department of Physics, University of Maryland, Baltimore County, 1000 Hilltop Circle, Baltimore, MD 21250, USA

\item Institute of Theoretical Physics, Faculty of Mathematics and Physics, Charles University, Prague, Czech Republic

\item Department of Astronomy and Theoretical Astrophysics Center, University of California, Berkeley, CA 94720, US

\item Department of Physics and Space Science, Royal Military College of Canada, PO Box 17000, Station Forces, Kingston, ON K7K 7B4, Canada

\item Institute of Communication Networks and Satellite Communications, Graz University of Technology, Infeldgasse 12, 8010 Graz, Austria

\item D\'{e}pt. de physique, Univ. De Montr\'{e}al, C.P. 6128, Succ. Centre-Ville, and Centre de Recherche en Astrophysique du Qu\'{e}ebec, Montr\'{e}eal, QC H3C 3J7, Canada

\item AAVSO, 49 Bay State Rd. Cambridge, MA 02138, USA

\item Instytut Astronomiczny, Uniwersytet Wroc{\l}awski, Kopernika 11, 51-622 Wroc{\l}aw, Poland

\item Silesian University of Technology, Institute of Electronics, Akademicka 16, 44-100 Gliwice, Poland

\item Institute for Astrophysics, University of Vienna, Tuerkenschanzstrasse 17, A-1180 Vienna, Austria  

\item Universit\"{a}t Innsbruck, Institut f\" {u}r Astro- und Teilchenphysik, Technikerstrasse 25, A-6020 Innsbruck Austria

\item DARK, Niels Bohr Institute, University of Copenhagen, Lyngbyvej 2, DK-2100 Copenhagen {\O}, Denmark

\item School of Physical and Chemical Sciences, University of Canterbury, Private Bag 4800, Christchurch 8120, New Zealand

\item Nicolaus Copernicus Astronomical Center, Polish Academy of Sciences, Bartycka 18, PL 00–716 Warsaw, Poland

\item Department of Physics and Astronomy, Macquarie University, NSW 2109, Australia

\item School of Computing Engineering and Mathematics, Western Sydney University, Locked Bag 1797, Penrith, NSW 2751, Australia.

\item Departamento de F\'isica, Universidade Federal de Sergipe, Av. Marechal Rondon, S/N, 49000-000, S\~ao Crist\'ov\~ao, SE, Brazil

\item Observat\'orio Nacional, Rua Gal. Jos\'e Cristino 77, 20921-400, Rio de Janeiro, RJ, Brazil

\item Dept. of Physics \& Astronomy, Stony Brook University, Stony Brook, NY, USA.

\item Department of Astronomy, The Ohio State University, 140 West 18th Avenue, Columbus, OH 43210, USA

\item South African Astronomical Observatory, P.O. Box 9, 7935 Observatory, South Africa

\item School  of  Physics and Monash Centre for Astrophysic,  Monash  University, Clayton,  Victoria3800, Australia

\item Astrophysics Research Institute, Liverpool John Moores Univ., Liverpool, L3 5RF, UK 

\item Southern African Large Telescope Foundation, PO Box 9, Observatory 7935, South Africa

\item Sternberg Astronomical Institute, Lomonosov Moscow State University, Universitetskii pr. 13, Moscow, 119992 Russia

\item Department of Physics and Astronomy, West Virginia University, P.O. Box 6315, Morgantown, WV 26506, USA

\item Center for Gravitational Waves and Cosmology, West Virginia University, Chestnut Ridge Research Building, Morgantown, WV 26505, USA

\item National Radio Astronomy Observatory, P.O. Box O, Socorro, NM 87801, USA

\item INAF-Osservatorio Astronomico di Trieste, Via G.B. Tiepolo 11, I-34143 Trieste, Italy

\item INAF--Osservatorio di Padova, vicolo dell’ Osservatorio 5, I-35122 Padova, Italy

\item Department of Astronomy, University of Wisconsin, 475 N. Charter Str., Madison, WI 53704, USA

\item School of Physics \& Astronomy, University of Leicester, LE17RH, UK

\item Institute for Astronomy, University of Hawai'i, 2680 Woodlawn Drive, Honolulu, HI 96822, USA

\end{affiliations}
\newpage


\begin{abstract}

Classical novae are thermonuclear explosions that occur on the surfaces of white dwarf stars in interacting binary systems\cite{Bode_etal_2008_mt}. It has long been thought that the luminosity of classical novae is powered by continued nuclear burning on the surface of the white dwarf after the initial runaway\cite{Gallaher_etal_1978_mt}. However, recent observations of GeV $\gamma$-rays from classical novae have hinted that shocks internal to the nova ejecta may dominate the nova emission. Shocks have also been suggested to power the luminosity of events as diverse as stellar mergers\cite{Metzger_Pejcha_2017_mt}, supernovae\cite{Moriya_etal_2018_mt}, and tidal disruption events\cite{Roth_etal_2016_mt}, but observational confirmation has been lacking. 
Here we report simultaneous space-based optical and $\gamma$-ray observations of the 2018 nova V906 Carinae (ASASSN-18fv), revealing a remarkable series of distinct correlated flares in both bands. The optical and $\gamma$-ray flares occur simultaneously, implying a common origin in shocks. During the flares, the nova luminosity doubles, implying that the bulk of the luminosity is shock-powered. Furthermore, we detect concurrent but weak X-ray emission from deeply embedded shocks, confirming that the shock power does not appear in the X-ray band and supporting its emergence at longer wavelengths. Our data, spanning the spectrum from radio to $\gamma$-ray, provide direct evidence that shocks can power substantial luminosity in classical novae and other optical transients. 

\end{abstract}

\beginmain

In a classical nova, the accreted envelope\cite{Bode_etal_2008_mt} (mass $\approx 10^{-7}-10^{-3}$ M$_{\odot}$) expands and is ejected at velocities of $\sim$500--5000\,km\,s$^{-1}$. The result is an optical transient where the luminosity of the system increases by a factor of $\sim 10^3-10^6$, sometimes making the source visible to the naked eye\cite{Warner_1995_mt}. After the initial ejection of the envelope, residual nuclear burning continues on the surface of the hot white dwarf, leading to a phase of quasi-constant, near-Eddington luminosity powered by the hot white dwarf\cite{Gallaher_etal_1978_mt,Wolf_etal_2013_mt}.
This should manifest as an optical light curve smoothly declining from maximum light, as the photosphere recedes and the peak of the spectral energy distribution moves blueward from the optical into the ultraviolet and finally into soft X-ray\cite{Bode_etal_2008_mt}. 
However, some novae show erratic flares around maximum light with a variety of timescales and amplitudes\cite{Strope_etal_2010_mt}; these features are still poorly explored and their origin remains a matter of debate. Proposed explanations include instabilities in the envelope of the white dwarf leading to multiple ejection episodes\cite{Cassatella_etal_2004_mt,Hillman_etal_2014_mt},
instabilities in an accretion disk that survived the eruption\cite{Goranskij_etal_2007_mt}, and variations in mass transfer from the secondary to the white dwarf\cite{Chocol_Pribulla_1998_mt}.

\begin{figure}
\begin{center}
\includegraphics[width=0.86\columnwidth]{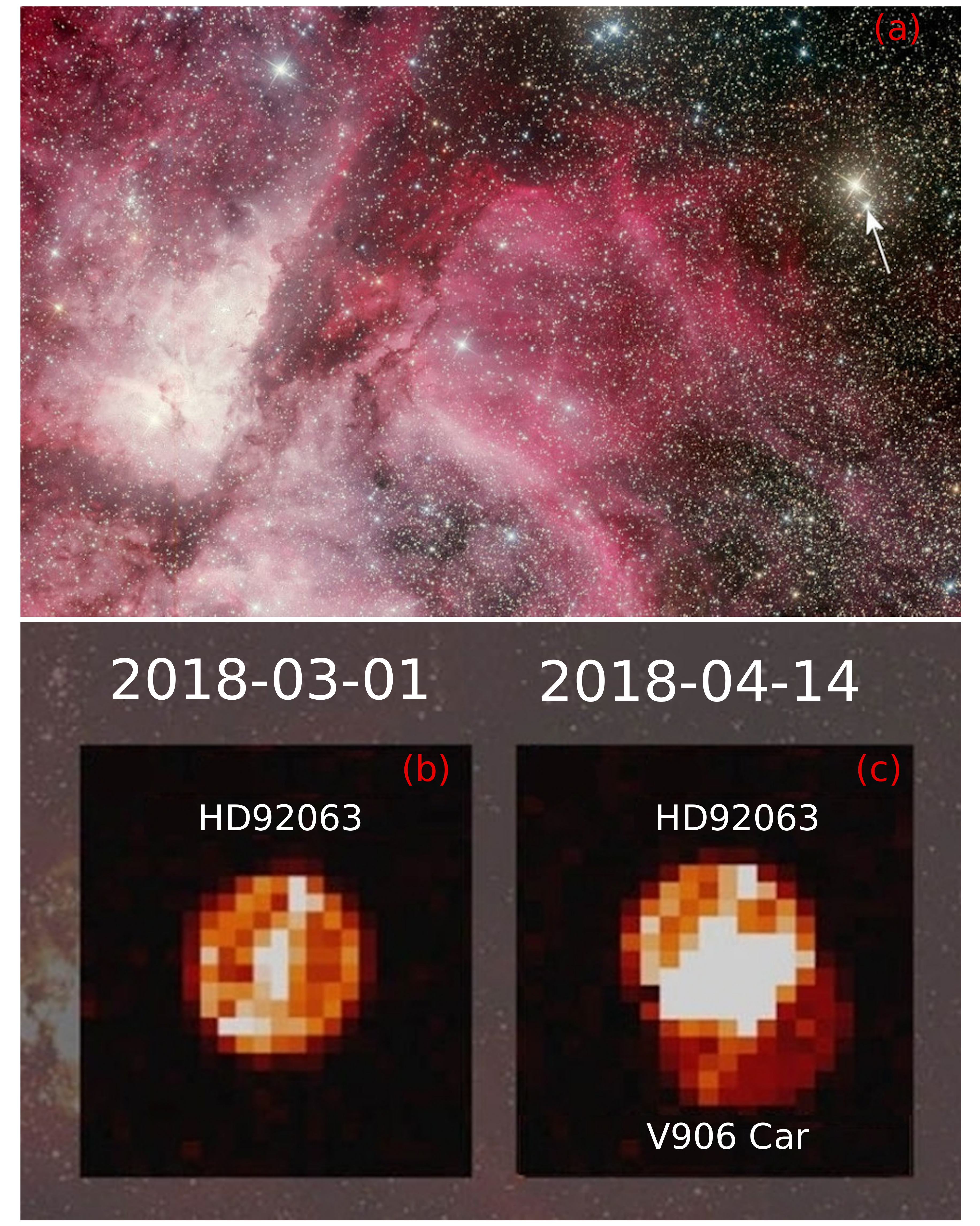}
\caption{\textbf{Nova V906~Car was discovered in a complex region of the Galaxy near the Carina nebula and the red-giant star HD~92063, which was being monitored by the BRITE satellite constellation.}
\textit{Panel (a):} V906~Car pictured in outburst and marked by a white arrow. HD~92063 is the brighter nearby star (Image credits: A. Maury \& J. Fabrega).
\textit{Panel (b):} BRITE image from before the outburst of V906~Car taken on 2018 March 1, showing only HD~29063. \textit{Panel (c):} BRITE image during the outburst taken on 2018 April 14, showing both HD~92063 and V906~Car.}
\label{Fig:1}
\end{center}
\end{figure}

\begin{figure}
\begin{center}
  \includegraphics[width=\textwidth]{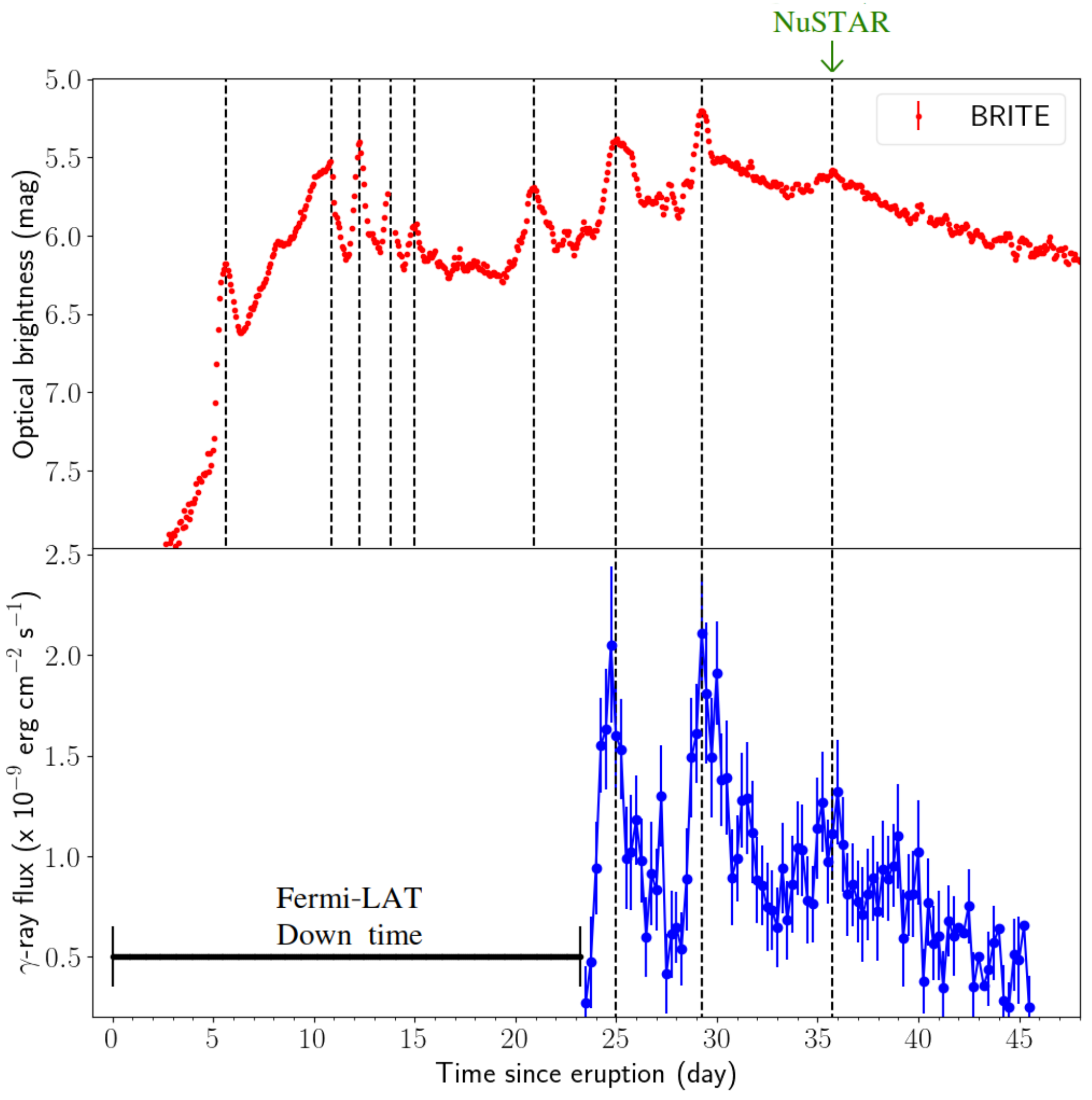}
\caption{\textbf{The optical and GeV $\gamma$-ray light curves of Nova V906~Car are correlated, showing simultaneous flares in both bands.} 
 The black dashed lines represent the dates of the post-maximum flares. The green arrow indicates the date of the first \textit{NuSTAR} X-ray observation. The black solid bar indicates the period of \textit{Fermi}/LAT down time due to technical issues. \textit{Fermi} entered another observing gap between days 46 and 57. The error bars in the BRITE light curve are 1$\sigma$ uncertainties. The point-to-point scatter of the binned BRITE measurements is $\approx$ 2\,mmag and therefore the size of the error bars is smaller than the symbols size. The error bars in the \textit{Fermi} light curve are 1$\sigma$ uncertainties. The eruption start is on 2018 March 16.03 UT (see \textit{Methods} for more details).}
\label{Fig:2}
\end{center}
\end{figure}

The optical transient V906~Carinae (ASASSN-18fv) was discovered by the All-Sky Automated Survey for Supernovae (ASAS-SN\cite{Shappee_etal_2014_mt}) on 2018 March 20.3 UT, and was shortly thereafter spectroscopically confirmed as a classical nova\cite{ATel_11454_mt,ATel_11460_mt}.
Serendipitously, V906~Car happened to occur in a field being monitored by the 
BRight Target Explorer (BRITE) nanosatellite constellation\cite{Pablo_etal_2016_mt} (Figure~\ref{Fig:1}), resulting in a high cadence optical light curve tracking the evolution of the eruption from its start (2018 March 16.13 UT; Figure~\ref{Fig:2}). The continuous, high cadence BRITE optical light curve (presented with 1.6\,hr resolution in Figure \ref{Fig:2}, the orbital period of the satellite) revealed a series of eight post-maximum flares during the first month of the outburst, each
lasting $\sim$ 1\,--\,3 days with amplitudes $\lesssim$ 0.8\,mag (Figure~\ref{Fig:2}; for more details see \textit{Methods} and Supplementary Information.\ref{SI_1}, hereafter SI). Typically, novae are observed using ground-based instruments at lower cadence, and light curves often contain substantial gaps, implying that such short timescale variability would be difficult to resolve. 

V906~Car was detected in GeV $\gamma$-rays around 23 days after eruption by the Large Area Telescope (LAT) on the \emph{Fermi Gamma-Ray Space Telescope}. The $\gamma$-rays persisted at least until day 46 after eruption\cite{ATel_11546_mt} (Figure~\ref{Fig:2}). The start time of the $\gamma$-ray emission is unconstrained, as the LAT was offline during the first 23 days of the eruption. The GeV $\gamma$-ray flux reached $2.1 \times 10^{-9}$\,erg\,cm$^{-2}$\,s$^{-1}$ on days 25 and 29, making V906~Car the brightest $\gamma$-ray nova to date\cite{Franckowiak_etal_2018_mt, Li_etal_2017_nature_mt}. Current theory suggests that the GeV $\gamma$-rays originate from shocks internal to the nova ejecta---specifically as a fast biconical wind slams into a slower equatorial torus\cite{Chomiuk_etal_2014_mt, Metzger_etal_2015_mt}. The shocks accelerate particles to relativistic speeds and $\gamma$-rays are produced when these relativistic particles interact with either the surrounding medium or seed photons\cite{Tatischeff_Hernanz_2007_mt,Martin_etal_2018_mt}.

The exceptional $\gamma$-ray brightness of V906~Car allowed us to obtain the most detailed $\gamma$-ray light curve of a nova to date, showing multiple $\gamma$-ray peaks. Comparing this with the BRITE light curve, we see that the $\gamma$-ray peaks coincide in time with the optical flares (Figure~\ref{Fig:2}). This correlation implies that the optical and $\gamma$-ray emission in novae share a common origin\cite{Li_etal_2017_nature_mt}. One possibility is that the luminosity in both bands is driven by shock power---much as has been theorized to occur in Type~IIn supernovae\cite{Chugai_etal_2004_mt}.
The typical expansion velocities of nova ejecta ($\sim$1,000\,km\,s$^{-1}$) and timing of the $\gamma$-rays ($\sim$weeks after outburst) imply that the shocked material must have high densities\cite{Metzger_etal_2015_mt} ($\sim 10^{10}$\,cm$^{-3}$). 
At these densities, shocks are expected to be radiative\cite{Metzger_etal_2015_mt} (i.e., the bulk of the shock energy emerges as radiation). 
Shocks of $\gtrsim$1,000\,km\,s$^{-1}$ heat gas to $\gtrsim 10^7$ K, and therefore typically emit thermal X-rays\cite{2015SSRv..188..187S_mt}.
However, at the high densities in nova shocks, the X-ray emission is likely attenuated and/or reprocessed into lower energy radiation, possibly due to a combination of efficient absorption\cite{Metzger_etal_2015_mt} and X-ray suppression in corrugated shock fronts\cite{Steinberg_Metzger_2018_mt}---and therefore the bulk of the shock luminosity may emerge as optical/infrared light.

The observed $\gamma$-ray luminosity, $L_{\gamma} \approx {\rm few} \times 10^{36}\ (d/4\,{\rm kpc})^2$ erg\,s$^{-1}$ in V906~Car, 
implies an energetic shock (we assume a distance $d$ = $4.0\pm1.5$\,kpc to the nova; see SI.\ref{SI_1}). Typically, only a few percent of the shock energy goes into the acceleration of relativistic particles\cite{Caprioli_Spitkobsky_2014_mt}, and $\sim20\%$ of this energy is emitted in the \emph{Fermi}-LAT pass-band\cite{Metzger_etal_2015_mt}. Therefore, the kinetic power of the shock is required to be $\gtrsim 10^{38}$\,erg\,s$^{-1}$---implying that the shock luminosity in V906~Car rivals the bolometric luminosity of the nova ($\sim$ a few 10$^{38}$\,erg\,s$^{-1}$; see SI.\ref{SI_5}) and likely outstrips the radiative luminosity from the nuclear-burning white dwarf ($\sim 10^{38}$\,erg\,s$^{-1}$, which is $\sim$ the Eddington Luminosity L$_{\mathrm{Edd}}$ for a 1\,M$_{\odot}$ white dwarf). 
Meanwhile, the optical--$\gamma$-ray correlation implies that the two wavebands share a common source---shocks---and therefore that shock luminosity is emerging in the optical band. This challenges the standard paradigm which attributes the bolometric luminosity of novae to thermal energy from the white dwarf. 

We can test where in the electromagnetic spectrum the shock luminosity emerges using X-ray observations concurrent with the $\gamma$-ray detections. Softer X-rays ($<10$ keV) are usually not detected while novae are observed to emit $\gamma$-rays\cite{Metzger_etal_2015_mt,Nelson_etal_2019_mt}, and V906~Car is no exception. The X-ray Telescope (XRT) on the \textit{Neil Gehrels Swift Observatory} monitored V906~Car during the $\gamma$-ray emission on days 5 and 37, but no X-rays were detected in the 0.3--10.0\,keV range with a 3$\sigma$ upper limit on the observed luminosity, $L_X < 4 \times 10^{33}\ (d/4\, \mathrm{kpc})^2 $\,erg\,s$^{-1}$.
However, these observations cannot rule out the presence of luminous, but heavily absorbed, X-ray emission. On day 36, coinciding with the last optical/$\gamma$-ray flare, we detected harder (3.5--78.0\,keV) X-rays from V906~Car with the \emph{NuSTAR} satellite (see \textit{Methods}). The detected X-rays were consistent with a highly-absorbed ($N_H = 1.9 \times 10^{23}$\,cm$^{-2}$) thermal plasma with an unabsorbed luminosity of $L_{\rm X} = (2.4 \pm 0.2) \times 10^{34}\ (d/4\, \mathrm{kpc})^2$\,erg\,s$^{-1}$ (see SI.\ref{SI_5}).
Therefore, the GeV $\gamma$-ray luminosity of V906~Car is a factor of $\sim$300 higher than the hard X-ray luminosity. This $L_X/L_{\gamma}$ is consistent with theoretical predictions of heavy absorption and X-ray suppression in nova shocks\cite{Steinberg_Metzger_2018_mt}, and indicates that the majority of the shock luminosity is indeed emitted in the optical band.

An alternative explanation for the optical--$\gamma$-ray correlation in V906~Car is if the particle acceleration were very efficient ($>10\%$) and variations in luminosity of the binary (perhaps on the white dwarf surface or in the accretion disk) power the flares in the optical light curve. These luminosity variation would lead to time-variable, radiation-driven outflows which in turn produce shocks and $\gamma$-rays. 
In this case, the $\gamma$-ray flares should lag the optical flares in time (possibly by several days; see SI.\ref{SI_6}). The 
optical and $\gamma$-ray light curves of V906~Car enable us to test for any time lag between the optical and $\gamma$-ray emission---something that has never been possible for a nova previously. Our correlation analysis (see SI.\ref{SI_4}) implies that the $\gamma$-ray emission precedes the optical by $\sim 5.3 \pm 2.7$\,h (2$\sigma$ significance). 
We can also rule out that the $\gamma$-rays lag the optical by more than 2.5\,h at 3$\sigma$ significance. The optical and $\gamma$-ray light curves of V906~Car are thus inconsistent with thermal emission from the white dwarf, and provide strong evidence for shocks powering the optical flares.

Our high-resolution optical spectra (SI.\ref{SI_2}) and light curves, lead us to suggest the following scenario for V906~Car. A dense slowly-expanding torus (with an expansion velocity $v_1 < 600$\,km\,s$^{-1}$) is ejected in the binary's orbital plane during the first days of the eruption\cite{Chomiuk_etal_2014_mt}. This torus has complex density structure, consisting perhaps of a spiral or multiple shells\cite{Pejcha_etal_2016_mt}.
Later, a fast wind develops with expansion velocity, $v_2 \approx 1200$\,km\,s$^{-1}$, and shocks the torus, presumably leading to $\gamma$-rays and other shock-powered emission.
The first four optical flares 
are created when the wind slams into the higher density structures in the torus, leading to temporary increases in the output shock luminosity.
Around 20 days after the eruption, we witness an even faster wind emerging ($v_3 \approx 2500$\,km\,s$^{-1}$; see SI.\ref{SI_2}). The faster wind slams into the torus, which is now merged with the previous wind, leading to the second sequence of the optical and correlated $\gamma$-ray flares (between 22 and 36 days after eruption). This multiple-ejection scenario is supported by our radio light curve of V906~Car, which is consistent with a delayed expansion of the bulk of the ejecta ($\sim$10--20 days after eruption) and is not consistent with a single homologous ejection (see SI.\ref{SI_3}).
While this scenario might be unique to V906~Car given the diversity of nova observations\cite{Strope_etal_2010_mt}, it supports a general unified picture in which a substantial fraction of nova emission is radiated by internal shocks. The correlation also suggests that the long debated flares seen in the optical light curves of some novae are originating from shock interactions.

The timescales and luminosities of other optical transients, such as Type IIn, Ia-CSM, and super-luminous supernovae, have led to the conclusion that these events are shock-powered---that is, the bulk of their bolometric luminosity originates as X-rays from shocks that are then absorbed and reprocessed to emerge in the optical\cite{2007ApJ...671L..17S_mt,2013ApJS..207....3S_mt,Moriya_etal_2018_mt}. Similar suggestions have been made for luminous red novae\cite{Metzger_Pejcha_2017_mt}, stellar mergers, and tidal disruption events\cite{Roth_etal_2016_mt}. Shocks are often theorized as a flexible way to power the most luminous transients in the sky\cite{Dong_etal_2016_mt,Chatzopoulos_etal_2016_mt}. However, there has never been direct evidence for shocks dominating the bulk of the bolometric luminosity of transients. Our observations of nova V906~Car definitively demonstrate that substantial luminosity can be produced---and emerge at optical wavelengths---by heavily-absorbed, energetic shocks in explosive transients. They also show that these same shocks can accelerate charged particles to relativistic speeds, implying that shock-powered supernovae may be important sources of cosmic rays\cite{Murase_etal_2014_mt,Murase_etal_2019_mt}. With modern time-domain surveys like ASAS-SN, the Zwicky Transient Facility (ZTF), and the Vera C. Rubin Observatory we will be discovering more---and higher luminosity---transients than ever before. The novae in our Galactic backyard will remain critical for testing the physical drivers powering these distant, exotic events.


\begin{addendum}

\item[Author Contributions] E.Ay.\ wrote the text. A.Pi., A.Po., R.Ku., K.V.So., L.Ch., S.Ry., M.Fi., R.Al., P.Ma., R.L.Ol., J.St., K.L.Li., A.Ki., L.Iz., F.M.W., and K.R.P.\ obtained and reduced the data. All authors contributed to the interpretation of the data and commented on the final manuscript.

\item[Competing Interests] The authors declare that they have no competing financial interests.

\item[Correspondence] Correspondence and requests for materials should be addressed to E.A. (email: aydielia@msu.edu), K.V.S. (email: kirx@kirx.net), and L.C. (email: chomiuk@msu.edu).

\item E.A., L.C., and K.V.S.\ acknowledge NSF award AST-1751874, NASA award 11-Fermi 80NSSC18K1746, and a Cottrell fellowship of the Research Corporation. K.L.L.\ was supported by the Ministry of Science and Technology of Taiwan through grant 108-2112-M-007-025-MY3. J.S.\ was supported by the Packard Foundation. O.P.\ was supported by Horizon 2020 ERC Starting Grant ``Cat-In-hAT'' (grant agreement \#803158) and INTER-EXCELLENCE grant LTAUSA18093 from the Czech Ministry of Education, Youth, and Sports. Support for K.J.S.\ was provided by NASA through the Astrophysics Theory Program (NNX17AG28G). G.A.W.\ acknowledges Discovery Grant support from the Natural Sciences and Engineering Research Council (NSERC) of Canada. A.F.J.M.\ is grateful for financial assistance from NSERC (Canada) and FQRNT (Quebec). A.Pi.\ acknowledges support provided by the Polish National Science Center (NCN) grant No.\ 2016/21/B/ST9/01126. A.Po.\ was supported by statutory activities grant SUT 02/010/BKM19 t.20. D.A.H.B.\ gratefully acknowledge the receipt of research grants from the National Research Foundation (NRF) of South Africa. A.Kn.\ acknowledges the National Research Foundation of South Africa and the Russian Science Foundation (project no.14-50-00043). RK, WW and KZ acknowledge support from the Austrian Space Application Programme (ASAP) of the Austrian Research Promotion Agency (FFG). I.V acknowledges the support by the Estonian Research Council grants IUT26-2 and IUT40-2, and by the European Regional Development Fund (TK133). This research has been partly founded by the National Science Centre, Poland, through grant OPUS 2017/27/B/ST9/01940 to J.M. This work is based on data collected by the BRITE Constellation satellite mission, designed, built, launched, operated and supported by the Austrian Research Promotion Agency (FFG), the University of Vienna, the Technical University of Graz, the University of Innsbruck, the Canadian Space Agency (CSA), the University of Toronto Institute for Aerospace Studies (UTIAS), the Foundation for Polish Science \& Technology (FNiTP MNiSW), and National Science Centre (NCN). GH is indebeted to the Polish National Science Center for funding by grant No. 2015/18/A/ST9/00578. C.S.K.\ is supported by NSF grants AST-1908952 and AST-1814440. We acknowledge the use of public data from the Swift data archive. UK funding for the Neil Gehrels Swift Observatory is provided by the UK Space Agency. 
This research has made use of data and/or software provided by the High Energy Astrophysics Science Archive Research Center (HEASARC), which is a service of the Astrophysics Science Division at NASA/GSFC and the High Energy Astrophysics Division of the Smithsonian Astrophysical Observatory. LI was supported by grants from VILLUM FONDEN (project number 16599 and 25501). A part of this work is based on observations made with the Southern African Large Telescope (SALT), under the Large science Programme on transient 2018-2-LSP-001. Polish participation in SALT is funded by grant No.\ MNiSW DIR/WK/2016/07. 
The Australia Telescope Compact Array is part of the Australia Telescope National Facility which is funded by the Australian Government for operation as a National Facility managed by CSIRO.
We acknowledge ARAS observers Terry Bohlsen, Bernard Heathcote, Paul Luckas for their optical spectroscopic observations which complement our database. Nova research at Stony Brook is supported in part by NSF grant AST 1614113, and by research support from Stony Brook University. We thank Encarni Romero Colmenero for initiating the collaboration that has led to this paper. 

\end{addendum}



\clearpage

\clearpage


\clearpage
\begin{center}
{\bf {\Large Methods}}
\end{center}
\subsection{The ASAS-SN discovery.}
V906 Car (ASASSN-18fv) was first discovered as a possible, bright Galactic nova by the All-Sky Automated Survey for Supernovae (ASAS-SN\cite{Shappee_etal_2014_mt,Kochanek_etal_2017_mt})
on 
2018 March 20.3 UT\cite{ATel_11454_mt} with $V <$ 10 mag (saturated). The nova is located at J2000 equatorial coordinates of $(\alpha, \delta)$ = (10$^{\mathrm{h}}$36$^{\mathrm{m}}$15$^{\mathrm{s}}$\!\!\!.\,42, --59$^{\circ}$35$'$54$''$\!\!\!.\,0) and Galactic coordinates of ($l, b$) = (286$^{\circ}$\!\!\!.\,580, --1$^{\circ}$\!\!\!.\,088; see Ref.\cite{ATel_11454_mt}). Shortly thereafter the transient was  confirmed spectroscopically as a classical nova\cite{ATel_11460_mt}. Pre-discovery observations obtained using Evryscope-South\cite{ATel_11467_mt}
 and BRITE  suggest that the eruption started on
2018 March 16.13 UT. Therefore we will assume this date as the eruption start ($t_0$).

\noindent \textbf{BRITE photometry.} 
On 2018 Feb 20, one of the five BRITE-Constellation\cite{Weiss_etal_2014_mt,Pablo_etal_2016_mt} satellites, BRITE-Toronto (BTr), started observations of 18 stars in the Carina field. Among those preselected objects was the red giant HD~92063 (K1 III, V=5.08 mag). The instrument aboard BTr is a five-lens, 3-cm-aperture telescope feeding an uncooled CCD. BTr is also equipped with a red filter transmitting light between 550 and 700\,nm. The pixel size of the BRITE CCD detector (KAI-11002M) is 9 $\mu$m and image scale is 27 arcsec/pixel.

The exposure time was set to 4 seconds and images were taken every 20 seconds. The observations were obtained over 16 minutes during each 98.24 min satellite orbit. 
Until 2018 March 18, no significant variability of HD~92063 was apparent in the BRITE photometry outside the 2\,mmag rms scatter. A few days later, on 2018 March 22, an upward trend in brightness was clearly noticeable.
An inspection of the images of HD~92063 revealed that another object appeared close to, in fact merging with, the point-spread function (PSF, FWHM $\approx$ 8 pixels or 3.6 arcmin) of the target (Figure~\ref{Fig:1}). This new source was discovered by ASAS-SN as ASASSN-18fv (V906 Car) and soon thereafter classified as a classical nova.


The observations of the nova were reduced with the standard BRITE pipeline\cite{Popowicz_etal_2017_mt}, which provides aperture photometry.
The raw BRITE photometry was subsequently processed to remove instrumental effects following the procedure outlined by Ref.\cite{Pigulski_2018_mt}. The procedure includes rejection of outliers and the worst orbits, and decorrelation of instrumental effect like CCD temperature and orbital phase. 

\noindent \textbf{$\gamma$-ray observations and analysis.}
The GeV $\gamma$-ray detection of V906~Car was first reported in Ref.\cite{ATel_11546_mt}. According to the preliminary report, $\gamma$-ray emission was significantly detected over the period 2018 April 14--18, but no detailed spectral or temporal information of the emission was provided. Here we analyze the \textit{Fermi}/LAT data to extract the $\gamma$-ray light curve and the spectral energy distribution (SED) of V906~Car. 

We downloaded the LAT data (Pass 8, Release 3, Version 2 with the instrument response functions of \texttt{P8R3\_SOURCE\_V2}) from the data server at the \textit{Fermi Science Support Center} (FSSC). The observations cover the period of 2018 April 8--30. We intended to use a broader time coverage; however, no LAT observations were performed for the region of interest (ROI) during March 17--April 7 and May 01--12 due to a solar panel issue on \textit{Fermi}. 

The data reduction and analyses were all performed using \texttt{fermitools} (version 1.0.5) with \texttt{fermitools-data} (version 0.17), which can be found at \url{https://fermi.gsfc.nasa.gov/ssc/data/analysis/software/}). For data selection, an ROI of $14^\circ\times14^\circ$ centred on the nova was used. 
Events with the class \texttt{evclass=128} (i.e., SOURCE class) and the type
\texttt{evtype=3} (i.e., reconstructed tracks FRONT and BACK) were selected. We excluded events with zenith angles larger than $90^\circ$ to avoid contamination from the Earth's limb. The selected events also had to be taken during good time intervals, which fulfils the \texttt{gtmktime} filter \texttt{(DATA\_QUAL$>$0)\&\&(LAT\_CONFIG==1)}. 

We then performed binned likelihood analysis on the selected LAT data. A $\gamma$-ray emission model for the whole ROI was built using all of the 4FGL cataloged sources\cite{2019arXiv190210045T_mt} located within $20^\circ$ of the nova. As V906~Car was the brightest $\gamma$-ray source in the field and its emission dominated within a 2$^\circ$ radius, we simply fixed all the spectral parameters of the field sources to the values in 4FGL to save computational time. 
In addition, the Galactic diffuse emission and the extragalactic isotropic diffuse emission were included by using the Pass 8 background models \texttt{gll\_iem\_v07.fits} and \texttt{iso\_P8R3\_SOURCE\_V2\_v1.txt}, respectively, which were allowed to vary during the fitting process.

Given the spectral curvature of V906~Car (see the SED in Supplementary Figure~\ref{Fig:Hard_Lept}), we considered two spectral models: a power law with an exponential cutoff, PLExpCutoff,
$$
\dfrac{dN}{dE} \propto E^{-\Gamma} \exp\Big(-\frac{E}{E_c}\Big),\\
$$
characterized by a photon index $\Gamma$ and energy cutoff $E_c$ (we find photon flux $F_{\mathrm{pht},\gamma}$ = 1.22 $\pm$ 0.04\,ph\,cm$^{-2}$\,s$^{-1}$, $E_c$ = 5.9$\pm$1.1\,GeV, and $\Gamma$ = 1.76 $\pm$ 0.05, with Test Statistic TS = 3591), where $N$ is the number of photons and $E$ is the photon energy; and a logarithmic power law, LogParabola:
\begin{equation}\nonumber
\dfrac{dN}{dE} \propto \Big(\frac{E}{E_b}\Big)^{-(\Gamma+\beta\log(E/E_b))},
\end{equation}
where $\beta$ defines the degree of curvature, and $E_b=100$~MeV is a fixed scale parameter. 
Both models can fit the data reasonably well. Formally, LogParabola performs better with $\Delta TS \approx 57$, so we used it as our standard model. 
The best-fit spectral parameters of the LogParabola fit to the full dataset are $\Gamma=1.14\pm0.09$, $\beta=0.21\pm0.02$, and $F\rm{_{ph} \,(0.1-300\,GeV)}=(1.04\pm0.04)\times10^{-6}$~photons cm$^{-2}$\,s$^{-1}$ with $TS=3649$.

We extracted the $\gamma$-ray light curve for 0.1--300~GeV using the LogParabola model with $\Gamma$ and $\beta$ fixed to their best-fit values (only the normalisation was allowed to vary). Several binning factors were tried and we finally used 6-hour bins as a good balance between the temporal resolution and the noise level of each bin (Figure~\ref{Fig:2}). The nova was significantly detected (greater than 5$\sigma$ significance) in most of the bins and 95\% upper limits were computed for bins with $TS<4$. 

We also extracted a 10-bin SED of V906~Car using geometric binning. To make the SED less model-dependent, we used a simple power-law model of $\Gamma=2$ (fixed; which is flat for a $\nu F_\nu$-based SED) to model the energy in each bin. As V906~Car is undetected in the last two bins of the SED (i.e., $TS\approx0$), we combined them to compute a 95\% upper limit. The final SED is shown in Supplementary Figure~\ref{Fig:Hard_Lept}. 

We fit a simple power law to daily-binned $\gamma$-ray data to search for any temporal variation in the energy spectrum. The photon index for the data-bins with TS$>$25 (i.e., significantly detected daily assuming a simple power-law) does not show significant changes. We conclude that there is no strong spectral variability in the data of nova V906~Car. The 1-day cadence light curve and the photon index of the 1-day binned data assuming a power-law fit can be found through this link: \url{http://scan.sai.msu.ru/~kirx/v906car/}

\noindent \textbf{X-ray observations with {\em Swift}.}
The X-ray Telescope (XRT) on the \textit{Neil Gehrels Swift Observatory} monitored V906~Car during the optical/$\gamma$-ray flaring. The observations obtained on 2018 March 21 and April 22, of 1ks and 1.25ks exposures, respectively, led to a non-detection in the 0.3--10.0\,keV X-ray range with a 3$\sigma$ upper limit on the observed luminosity, $L_X < 4.4 \times 10^{33}\ (d/4.0\, \mathrm{kpc})^2$\,erg\,s$^{-1}$. This implies an unabsorbed luminosity $L_X < 1.2 \times 10^{34} (d/4.0\mathrm{kpc})^2$\,erg\,s$^{-1}$, assuming a highly-absorbed thermal plasma. Supersoft (0.3--1\,keV) emission was only detected by \textit{Swift} $>$ 200 days after the eruption (Sokolovsky et al. 2020 in prep.).

\noindent \textbf{Hard X-ray observations with {\em NuSTAR} and {\em XMM-Newton}.}
Thanks to its unique optics capable of focusing X-rays in the energy range 3--78\,keV, {\em NuSTAR}\cite{2013ApJ...770..103H_mt} is two orders of magnitude more sensitive compared to the earlier coded aperture mask instruments operating at this energy range.  
V906~Car was observed with {\em NuSTAR} on 2018 April 21 and May 12 (36 and 57 days after the eruption). The nova was still bright in $\gamma$-rays during the first {\em NuSTAR} observation and no simultaneous {\em Fermi}/LAT observations are available during the second {\em NuSTAR} epoch. Both {\em NuSTAR} observations had integration times of about 48\,ks and resulted in highly significant detections (24$\sigma$ and 42$\sigma$, respectively) of the nova. The 3--78\,keV spectra lack obvious features like emission lines or absorption edges and can be fit with an absorbed thermal plasma model, with the temperature decreasing from $8.6 \pm 0.8$\,keV in the first epoch to $4.3 \pm 0.2$\,keV in the second epoch. To account for the lack of obvious spectral features in the {\em NuSTAR} band we have to assume non-solar metallicities (see below). 
Specifically, it is puzzling that the {\em NuSTAR} spectra show no signs of Fe emission and absorption while the optical spectra of the nova reveal strong Fe lines. An Fe abundance of less than $0.1$~Solar is required to fit the {\em NuSTAR} data. The alternative model that provides a good fit to our {\em NuSTAR} data has the Fe abundance fixed to the Solar value, but requires an overabundance of CNO elements by a factor of 200 over the Solar values. 
The ``CNO overabundance'' and ``Fe-defficiency'' models  imply a factor of 70 difference in the total absorbing column, while both models suggest values well in excess of the expected Galactic column. 

We performed a target of opportunity observation with {\em XMM-Newton} on 2018 December 16 (275 days post explosion) to constrain the abundances in the nova ejecta. The joint analysis of the high resolution X-ray grating (0.3--2.1\,keV) and medium-resolution CCD (0.3--10\,keV) {\em XMM-Newton} spectra suggests that the Fe abundance (by number) is less than 0.1 of the solar value, while the ejecta are enriched with N and O by factors of 350 and 30, respectively. Using these abundance values to fit the {\em NuSTAR} spectra 
we derive absorbing columns of $N_H = (1.9 \pm 1)\times10^{23}$\,cm$^{-2}$ on day 36, and $(2.6 \pm 0.2)\times10^{22}$\,cm$^{-2}$ on day 57. 
The unabsorbed fluxes (extrapolating to the 0.3--78\,keV energy range) are $1.3 \times 10^{-11}$\,erg cm$^{-2}$ s$^{-1}$ and $1.6 \times 10^{-11}$\,erg cm$^{-2}$ s$^{-1}$ (statistical uncertainty of $\sim20$\,\%) for the two epochs, respectively. A detailed discussion of the X-ray spectroscopy results will be presented by Sokolovsky et al.\ (2020, in prep.).
\\

\noindent \textbf{Data availability.} The data that support the plots within this paper and other findings of this study are available from \url{http://scan.sai.msu.ru/~kirx/v906car/} or from the corresponding authors upon reasonable request.



\beginsupplement

\clearpage
{\bf {\Large Supplementary Information}}

\section{Optical and Near-IR photometric observations and analysis}
\label{SI_1}

\subsection{SMARTS multi-band photometry.}
We make use of publicly available \textit{BVRIJHK} photometry \cite{Walter_etal_2012} from the ANDICAM instrument mounted on the 1.3-m telescope of the Small and Moderate Aperture Research Telescope System (SMARTS) at the Cerro Tololo Inter-American Observatory (CTIO) in Chile. The SMARTS data span 2018 March 22 to June 26 and offer simultaneous multi-band photometry, which is useful to estimate the bolometric luminosity of the nova and produce spectral energy distribution in the optical and near-infrared (NIR). The multi-band light curve is plotted in Supplementary Figure \ref{Fig:LC_BVRIJHK}.

\begin{figure}[!b]
\begin{center}
  \includegraphics[width=0.8\textwidth]{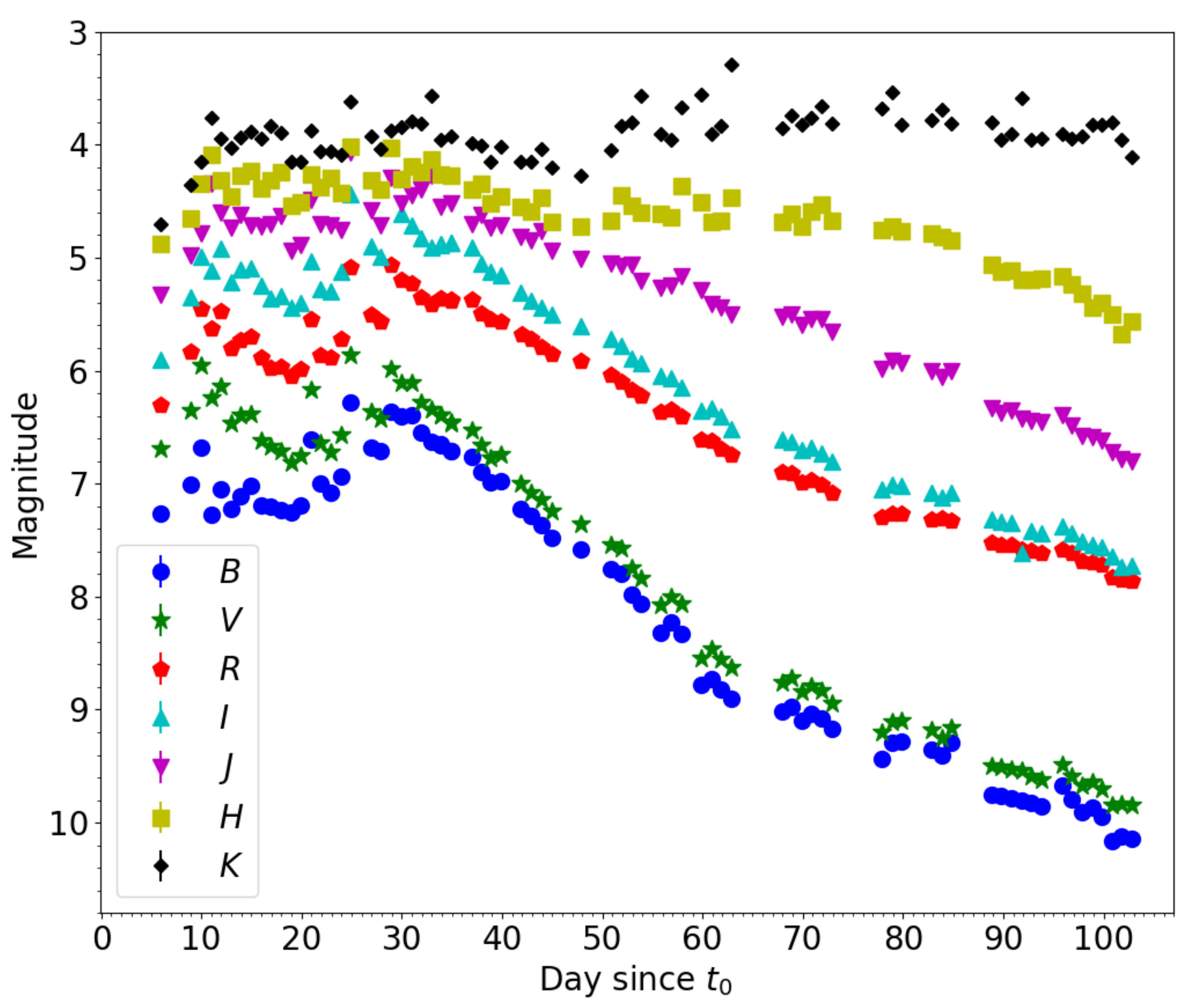}
\caption{\textbf{The SMARTS \textit{BVRIJHK} observations as a function of days since eruption.}}
\label{Fig:LC_BVRIJHK}
\end{center}
\end{figure}

\subsection{Optical light curve.} \label{opt_LC}
The most intriguing aspect of the optical light curve of V906~Car is the presence of multiple rapid flares superimposed on top of a more slowly rising and falling light curve.  
Here we derive the parameters typically used to classify nova light curves, such as the rise time to maximum, the maximum light, the decline time from maximum (the speed class), and the morphology of the decline. 

\begin{figure}[!t]
\begin{center}
  \includegraphics[width=0.9\textwidth]{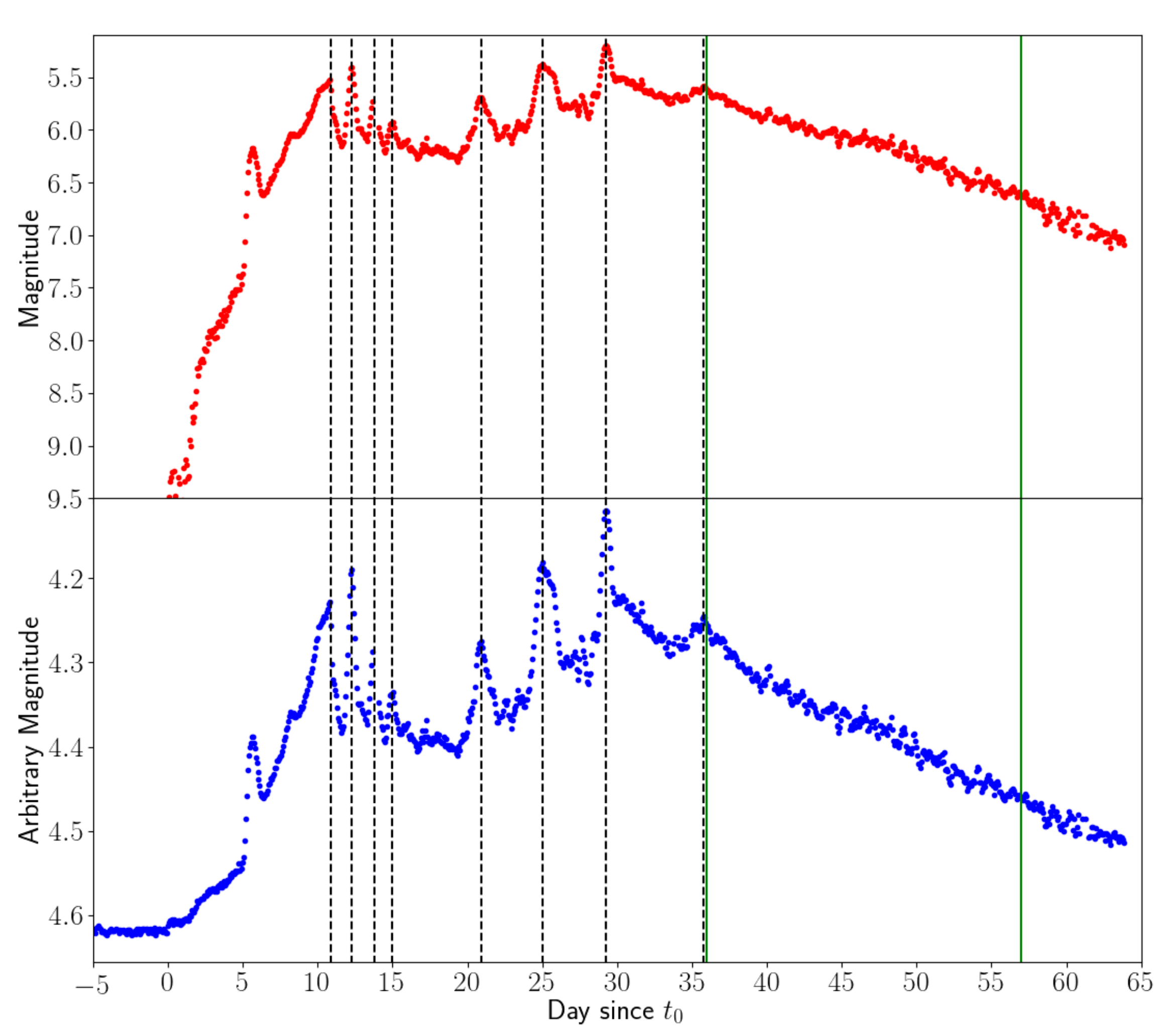}
\caption{\textbf{\textit{Top}: the complete high-cadence optical light curve of nova V906~Car from BRITE} reveals rapidly-evolving flares, during which the nova luminosity nearly doubles in $\sim$1 day. The vertical black dashed lines mark the dates of the peaks of optical flares. The green solid lines represent the dates of the \textit{NuSTAR} X-ray observations. The contribution of the red-giant star HD~92063 to the BRITE photometry was subtracted from the photometry assuming a constant magnitude of 4.62 in the BRITE band. \textit{Bottom}: the BRITE light curve without subtracting the contribution of the red giant star HD~92063 to show the onset of the rise.}
\label{Fig:sub}
\end{center}
\end{figure}

The BRITE optical light curve (Supplementary Figure~\ref{Fig:sub}) shows a moderately slow rise for 10.5 days from quiescence to the first maximum. During the rise, the light curve shows a flare-like brightness increase on day 5.6. This brightness bump resembles the pre-maximum halt observed in some novae (see figure 2.2 in Ref.\cite{Warner_2008}). 

\begin{figure}[!t]
\begin{center}
  \includegraphics[width=0.9\textwidth]{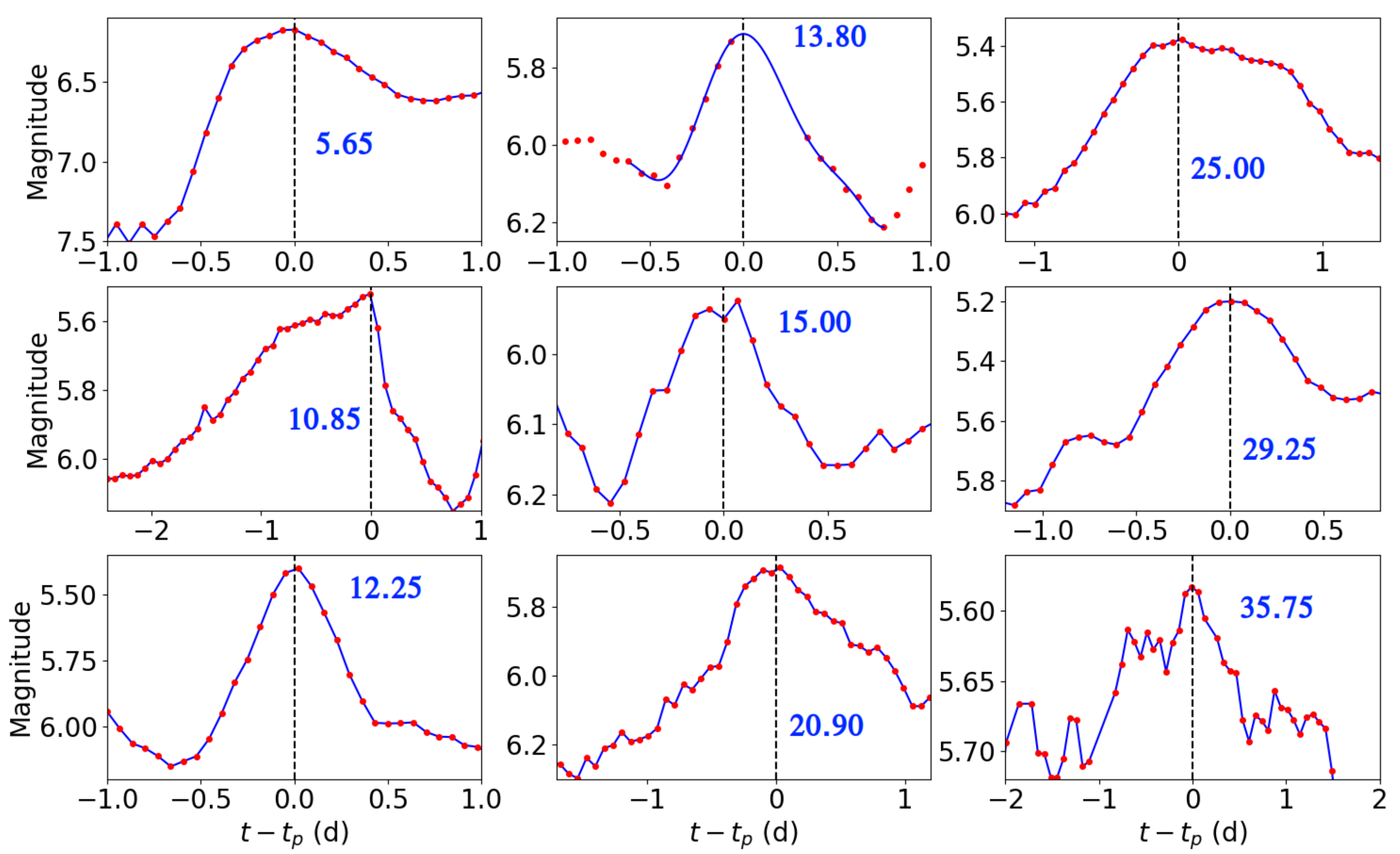}
\caption{\textbf{Profiles of each of the eight optical flares and the pre-maximum bump on day 5.65 from the BRITE data.} Magnitude is plotted against days from the peak of the flare ($t_p$), which is represented by the black dashed line; the number on each plot gives the time of the peak in units of days since t$_0$. For the flare on day 13.8 there is a gap in the observations, therefore we fit a high-order polynomial to the light curve to estimate the peak of the flare.}
\label{Fig:flares_profiles}
\end{center}
\end{figure}

\begin{table}[!t]
\centering
\caption{\textbf{The characteristics of the optical (\textit{top}) and $\gamma$-ray (\textit{bottom}) flares.} See text for more details.}
\begin{tabular}{rrrrrr}
\hline
Nr. & $t_{\mathrm{peak,opt}}$ & $\Delta t_{\rm opt}$ & $\Delta R$ & $dR/dt_{\rm rise}$ & $dR/dt_{\rm fall}$\\
 & (days) & (days) & (mag) & (mag\,d$^{-1}$) & (mag\,d$^{-1}$)\\
\hline
1 & 10.85 & 2.8 & 0.65 & 0.3 & 0.9\\
2 & 12.25 & 1.3 & 0.75 & 1.1 & 1.2\\
3 & 13.83 & 1.0 & 0.40 & 0.8 & 0.8\\
4 & 15.00 & 1.0 & 0.25 & 0.5 & 0.5\\
5 & 20.90 & 2.6 & 0.60 & 0.4 & 0.6\\
6 & 25.00 & 2.4 & 0.65 & 0.6 & 0.5\\
7 & 29.25 & 1.2 & 0.50 & 0.8 & 0.8\\
8 & 35.75 & 4.3 & 0.20 & 0.1 & 0.4\\
\hline
\end{tabular}
\begin{tabular}{rrrcc}
\hline
Nr. & $t_{\mathrm{peak},\gamma}$ & $\Delta t_{\gamma}$ & $F_{\mathrm{peak},\gamma}$ & $\Delta F_{\gamma}$\\ 
 & (days) & (days) & ($10^{-9}$\,erg\,cm$^{-2}$\,s$^{-1}$) & ($ 10^{-9}$\,erg\,cm$^{-2}$\,s$^{-1}$)\\
\hline
6 & 24.74 & 2.5 & 2.1 $\pm$ 0.4 & 1.8 $\pm$ 0.4\\
7 & 29.24 & 2.3 & 2.1 $\pm$ 0.3 & 1.6 $\pm$ 0.3\\
8 & 35.98 & 5.0 & 1.3 $\pm$ 0.3 & 0.7 $\pm$ 0.3 \\
\hline
\end{tabular}
\label{table:flares}
\end{table}

The light curve exhibits at least eight flares between days 9--36 after outburst (see Supplementary Figure \ref{Fig:sub} and Supplementary Table~\ref{table:flares}). These resemble the jitters of J-class nova light curves defined by Ref.\cite{Strope_etal_2010}. 
The flares are characterized by amplitudes of $\sim$0.3--0.8\,mag and they lasted for 1--3 days. After the first and second flares, the light curve shows an apparent decline in brightness and the amplitudes of the flares decrease (Supplementary Figure~\ref{Fig:sub}). The decline stops at day $\sim$20 post-eruption and both the brightnesses and amplitudes of the following flares increase. In the top portion of Supplementary Table~\ref{table:flares}, we characterize the time, amplitude, and rise/fall rate of the optical flares using the BRITE data. The time of the flare's maximum, in units of days after $t_0$, is $t_{\rm peak,opt}$, while $\Delta t_{\rm opt}$ is the duration between the start of the rise and the time when the brightness drops to the same level as before the rise (or the time when a new rise starts). $\Delta R$ is the difference between the magnitude at the rise and the magnitude at the peak of the flares, while the $dR/dt$ parameters give the rate of the magnitude change on the flare's rise and fall.
We find that the time interval between the peaks of the first four flares ($\lesssim$ 1.5\,days) is shorter than the time interval between the peaks of the last four flares ($\gtrsim$ 4\,days). However, we find no clear pattern in the amplitudes of the flares, nor their rise/decline rates, and the flares are asymmetric and have differing profiles (see Supplementary Figure~\ref{Fig:flares_profiles} for a zoom-in on the profiles of the flares). The BRITE light curve also shows low-amplitude variability or oscillations during the flaring period. These are particularly visible between flares, and they are characterized by an amplitude of $<$ 0.3\,mag and a timescale of $<$ 1 day.

V906~Car reached its ``first'' $V$-band brightness peak of $V \approx$ 5.94 on 2018 March 26.53 (day 10.5), implying that the rise rate from quiescence was $\sim$1.4\,mag per day. Yet, the nova reached its peak brightness on day 29.3 at $V_{\rm max}$ $\approx$ 5.86. Correcting for reddening ($A_V \approx$ 1.2; see SI.\ref{SI_2}), $V_\mathrm{max}$ implies $V_\mathrm{0, max} \approx 4.7 \pm 0.1$ mag. Using a distance of 4.0 $\pm$ 1.5\,kpc (see below), V906~Car's peak absolute magnitude is $M_{\rm V,max} \approx -8.3^{+1.3}_{-0.9}$ mag.

The time to decline from maximum by two magnitudes ($t_2$) is often used to classify nova light curves in terms of a speed class.   
We derive $t_2$ as the time for the brightness to decline by two magnitudes from both days 10.5 and 29.3, and find $t_2 = 44 \pm 1$\,d and $t_2 = 26 \pm 1$\,d in the $V$ band, respectively. This makes V906~Car a moderately fast nova\cite{Payne-Gaposchkin_1964,Warner_2008}. 

Past day 30, the light curve shows a rapid drop in the optical brightness accompanied by a rise in the near-IR flux from day $\sim$48 (Supplementary Figure~\ref{Fig:LC_BVRIJHK}). This signature in nova light curves is attributed to dust formation, as the dust grains absorb optical light and re-emit it in the IR\cite{Evans_Rawlings_2008}. During the decline, quasi-periodic oscillations on a timescale of $\sim$ 1\,d with an amplitude of $<$ 0.5\,mag appear, similar to O-class nova light curve classification of Ref.\cite{Strope_etal_2010}. Interestingly, similar oscillations appear simultaneously in the $\gamma$-ray light curve (Figure~\ref{Fig:2}). This indicates that the oscillations in the light curves also originate from shock interaction, possibly due to density inhomogeneities in the colliding ejecta.  

\begin{figure}[!t]
\begin{center}
  \includegraphics[width=0.9\textwidth]{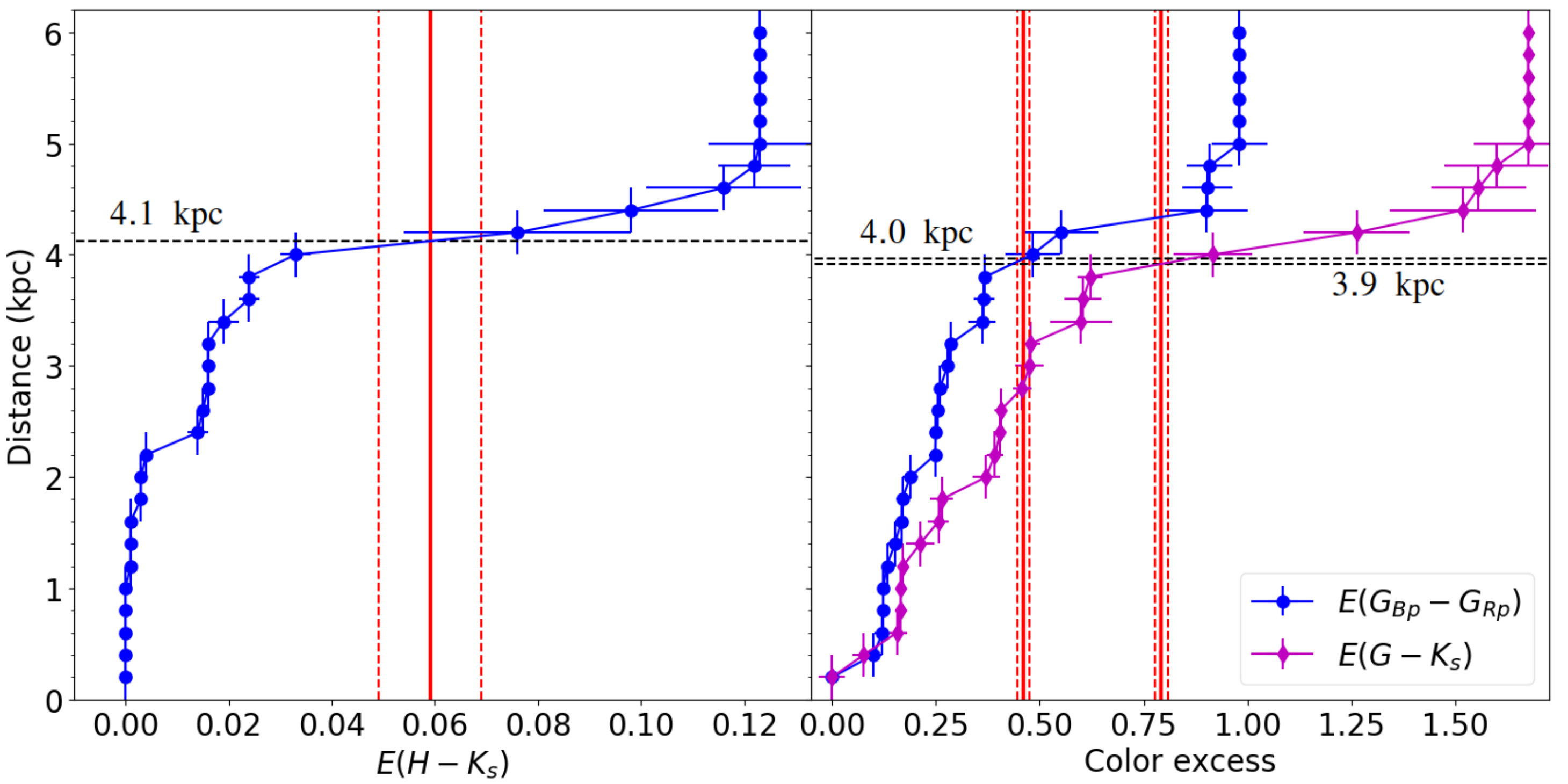}
\caption{\textbf{The distance-dependent extinction from the 3D Galactic reddening maps\cite{Chen_etal_2019} are used to constrain the distance to V906~Car.} \textit{Left:} the blue circles represent the $E(H-K_s)$ reddening as a function of distance at the position of V906~Car. The red solid bar is the derived reddening and the dashed red lines are the uncertainty on the derived value. The black dashed horizontal line represents the distance derived from the reddening maps and shows that the reddening indicates a distance $d \approx$ 4\,kpc. \textit{Right:} same as left, but for  $E(G_{Bp}-G_{Rp})$ (blue circles) and $E(G-K_s)$  (magenta circles). These reddening values also indicate a distance $d \approx$ 4\,kpc. The extinction errors are from the reddening measurements. The distance error bars are from the distribution of the Galactic maps. These error bars are 1$\sigma$ uncertainties.}
\label{Fig:dist_red}
\end{center}
\end{figure}

\subsection{Distance.}
V906~Car has a \emph{Gaia} DR2 counterpart (ID 5254540166225866496) with quiescent magnitude $G$ = 19.68 and parallax $P = 0.151 \pm 0.488$\,mas. Based on the method described in Ref.\cite{Bailer-Jones_etal_2018}, which uses a weak distance prior that varies smoothly as a function of Galactic longitude and latitude (by adopting a Galaxy model), we derive a distance range of 1.7--6.2\,kpc with a mode value of 3.3\,kpc. 
This distance should be considered uncertain due to the large error on the parallax and since \emph{Gaia} parallax measurements of a cataclysmic variable might be affected by brightness changes. Future \emph{Gaia} data releases will probably improve the parallax measurement and its associated uncertainty.

For an additional distance estimate we use 3D Galactic reddening maps\cite{Chen_etal_2019} combined with our reddening measurements from the spectral absorption features (see SI.\ref{SI_2}). The maps of Ref.\cite{Chen_etal_2019} use measurements from the \emph{Gaia} DR2, 2MASS and WISE surveys. Therefore, in order to make use of these maps, we convert our $E(B-V)$ measurement to reddening values in the 2MASS $JHK$ filters and the \emph{Gaia} DR2 $G$, $G_{Bp}$, and $G_{Rp}$ bands, using the extinction law from Refs.\cite{Chen_etal_2019, Wang_Chen_2019}. We find $E(H-K_s) = 0.06 \pm 0.01$, $E(G-K_s) = 0.79 \pm 0.02$, and $E(G_{RBp}-G_{Rp}) = 0.49 \pm 0.02$ for V906~Car. We use the closest Galactic coordinates in Ref.\cite{Chen_etal_2019} maps ($l$\,=\,286.550; $b =-1.050$) to estimate the distance towards the nova. In Supplementary Figure~\ref{Fig:dist_red} we present the reddening values as a function of distance in all three colours. Based on the maps we derive a distance to V906~Car of $4.1 \pm 0.2$\,kpc from $E(H-K_s)$, $4.0 \pm 0.2$\,kpc from $E(G_{RBp}-G_{Rp})$, and $3.9 \pm 0.2$\,kpc from $E(G-K_s)$. The average of the distances derived from the Galactic reddening maps is 4.0$\pm$ 0.2\,kpc, in good agreement with the Gaia measurements given the uncertainties. We adopt a distance of 4.0 $\pm$ 1.5\,kpc throughout the paper.

\subsection{Colour evolution and SEDs}
To better understand the changing conditions during the optical flares, we consider the colour evolution of V906~Car. 
In Supplementary Figure~\ref{Fig:colours}, we present the evolution of the reddening-corrected $(B - V)_0$, $(V - R)_0$, and $(R - I)_0$ colours. The colours show small, rapid changes correlated with some of the optical flares, but these changes are small compared to other flaring novae (see, e.g., Ref.\cite{Aydi_etal_2019_I}).



\begin{figure}
\begin{center}
  \includegraphics[width=\textwidth]{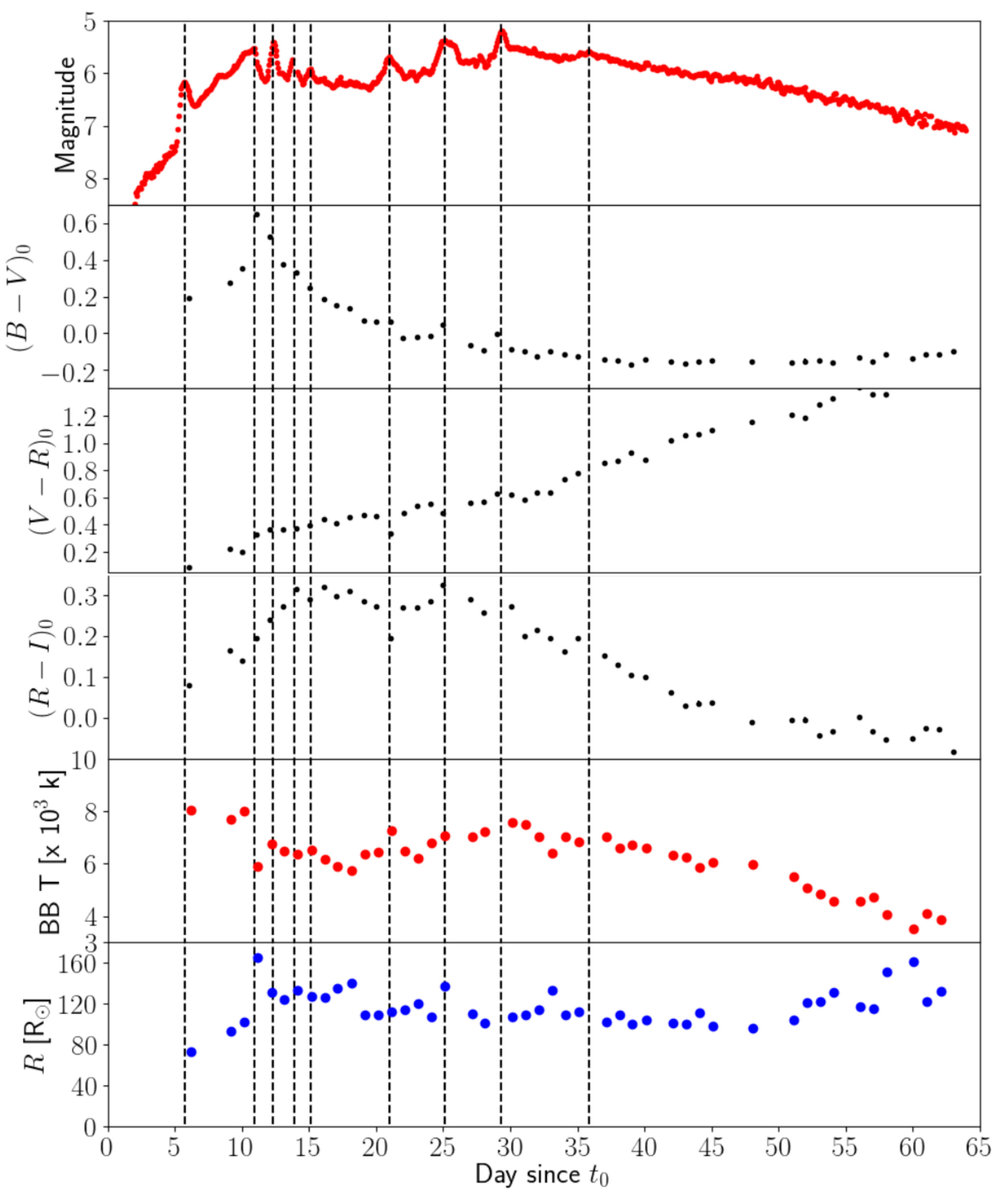}

\caption{\textit{From top to bottom}: \textbf{the BRITE optical light curve of V906~Car, the evolution of the optical broad-band colour $(B-V)_0$, $(V-R)_0$, and $(R-I)_0$, the evolution of the temperature and radius} derived from blackbody fitting (see Supplementary Figure~\ref{Fig:SEDs}). The errors on the colours are smaller than the symbols size.}
\label{Fig:colours}
\end{center}
\end{figure}

\begin{figure}[!t]
\begin{center}
  \includegraphics[width=0.85\textwidth]{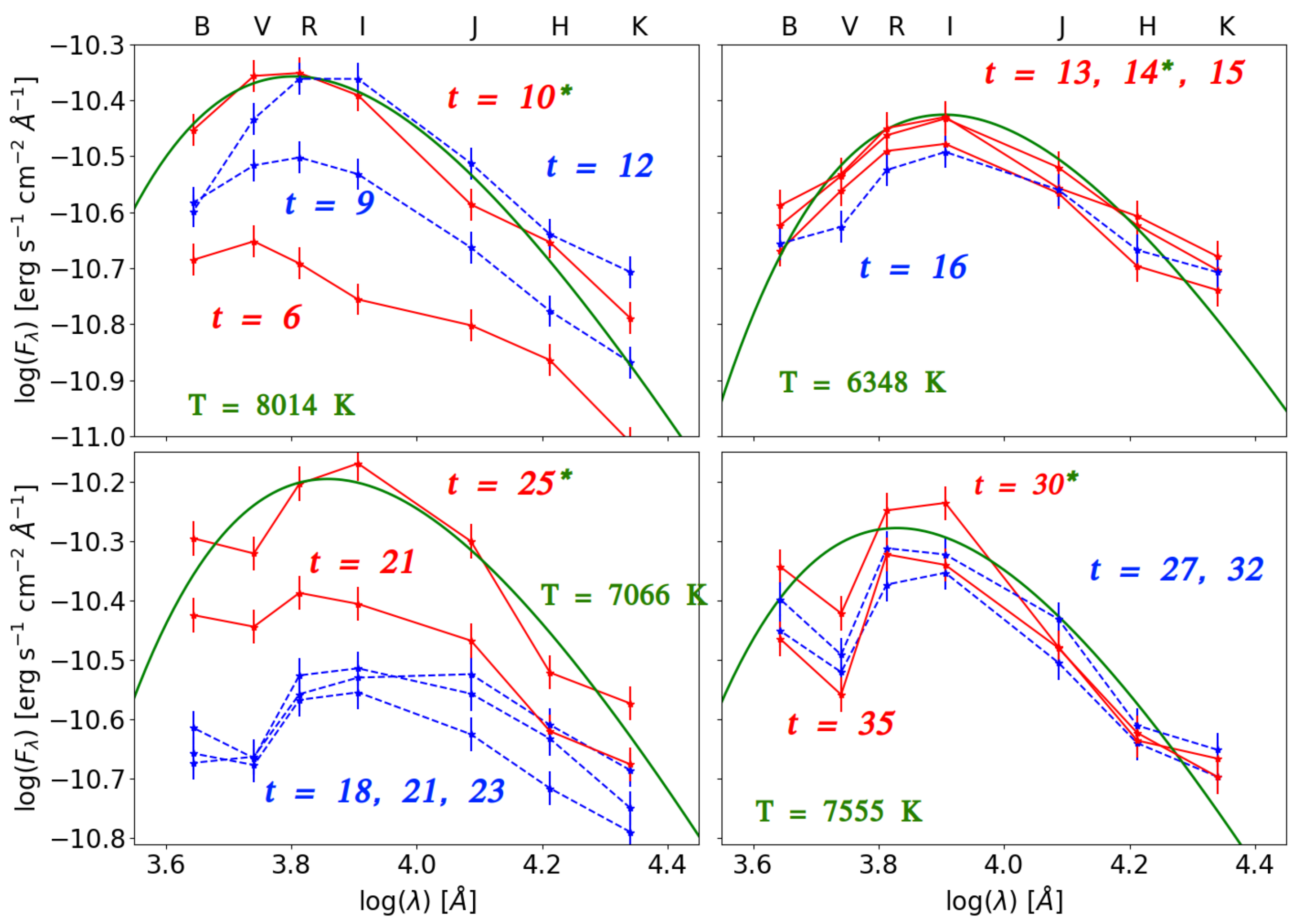}
\caption{\textbf{Extinction-corrected SED plots, showing the evolution of the SED during the first 35 days of the eruption of V906~Car.} The red SEDs are measurements obtained during a flare (maximum) and the blue ones are measurements obtained during a minimum. The error bars are 1$\sigma$ uncertainties and they include contributions from the photometric and extinction uncertainties. The green curves represents a sample of the best fit blackbody models. The days of the observations are indicated in red and blue with a green star indicating which SED is fitted with a blackbody curve on the plot. The temperatures quoted are derived from the best fit blackbody model.}
\label{Fig:SEDs}
\end{center}
\end{figure}

In order to study the evolution of the nova's photospheric temperature and radius, we produce spectral energy distributions (SEDs) spanning the  near-IR and optical bands. Typically, during optical maximum light, the photosphere of a nova reaches its maximum radius and the SED is characterized by an effective temperature in the range of 6000--8000\,K (e.g., Refs.\cite{Hachisu_Kato_2004,Bode_etal_2008}). During optical light curve decline, the photosphere shrinks in radius and the SED peaks in the UV and eventually in the soft X-rays.
For V906~Car during the first 15 days of its eruption, the SED is continuum dominated and can be well described as a blackbody peaking in the optical (Supplementary Figure~\ref{Fig:SEDs}). At later times, the emission lines start to dominate the SED, with two emission peaks in the $R$- and $B$-bands from the H Balmer and several \eal{Fe}{II} emission lines, particularly H$\alpha$ in the $R$ band and H$\beta$, H$\gamma$, and H$\delta$ in the $B$ band. In Supplementary Figure~\ref{Fig:colours} we present the evolution of the temperature derived from fitting a blackbody to each SMARTS multi-band epoch. No obvious correlation between the flares and the change of the photosphere temperature/radius can be inferred from Supplementary Figure~\ref{Fig:colours}.

\subsection{The secondary star and orbital period.}

The quiescent magnitude of the system $G \approx 19.7$ (which might be dominated by the accretion disk and the bright spot\cite{Warner_1995}), implies $G_0 \approx 18.8 \pm 0.1$, using $A_G = A_V \times 0.79$ from Ref.\cite{Wang_Chen_2019}. At an assumed distance of 4.0 $\pm$ 1.5\,kpc, the absolute quiescent magnitude of V906~Car would be $M_G = 5.8^{+1.3}_{-0.9}$. At this absolute magnitude, which should be considered as an upper limit due to the contribution of the disk and bright spot, the secondary star is an M dwarf. 

\begin{figure*}
\centering
  \includegraphics[width=0.48\textwidth]{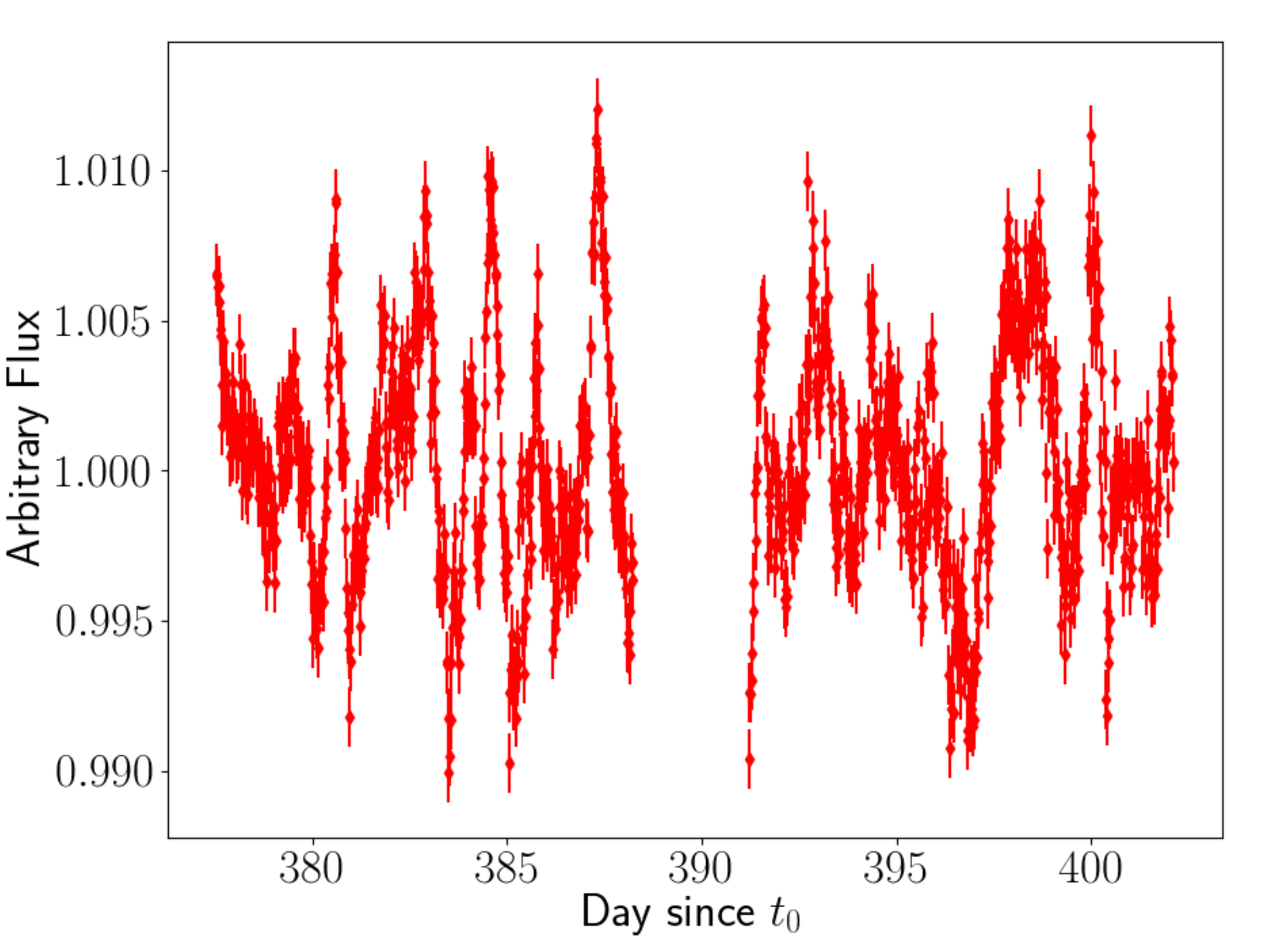}
    \includegraphics[width=0.5\textwidth]{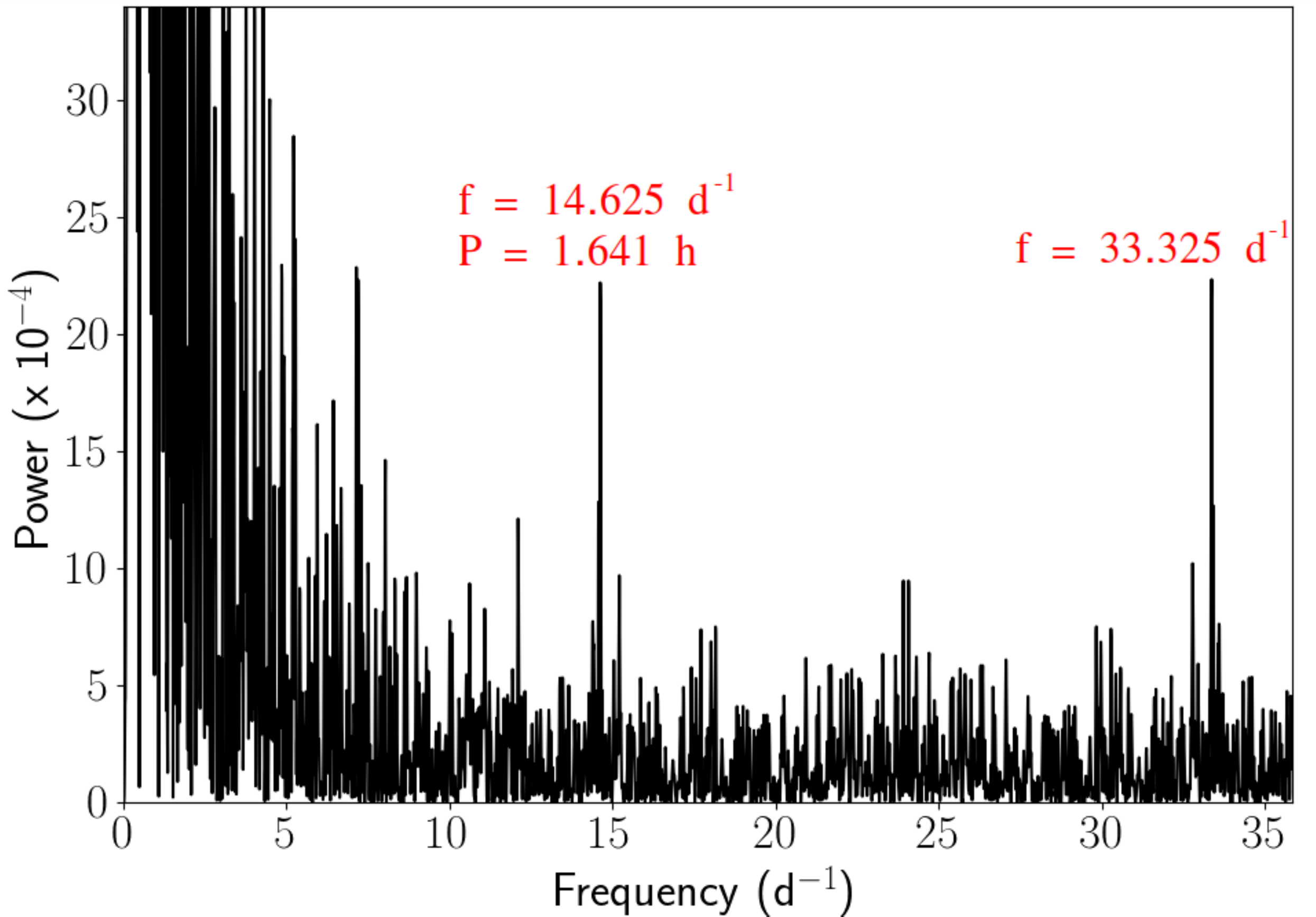}
\caption{\textbf{The TESS data and timing analysis for nova V906~Car}. \textit{Left:} the TESS light curve of V906~Car. The error bars are 1$\sigma$ uncertainties. \textit{Right:} the Lomb-Scargle periodogram of the de-trended TESS light curve, revealing modulation at a frequency of 14.625\,$d^{-1}$ (P $=$ 1.641\,h). The signal at 33.375\,$d^{-1}$ is an alias of this period.}
\label{Fig:TESS_LC}
\end{figure*}

The Transiting Exoplanet Survey Satellite (TESS\cite{Ricker_etal_2015}) obtained late-time observations of V906~Car during Sector 10 observations from 2019 March 26 to April 22. We used 
the open-source tool \textsc{eleanor}\cite{Feinstein_etal_2019} to extract light curves from the TESS full-frame images, opting to utilize the corrected flux light curve and to only include data that is not associated with a quality flag. This minimizes issues with background flux and artefacts introduced by the bright stars near V906~Car that dominate the raw flux light curve. 
The TESS light curve of V906~Car (Supplementary Figure~\ref{Fig:TESS_LC}) shows low-amplitude flickering with r.m.s. amplitude of 0.4\% (0.004 mag) on timescale of hours up to a day. 

We carried out timing analysis of the light curve using the Lomb-Scargle\cite{Scargle_1982} and Deeming\cite{Deeming_1975} methods. The Lomb-Scargle periodogram obtained from the detrended light curve shows strong modulation at lower frequencies which are likely due to red noise (Supplementary Figure~\ref{Fig:TESS_LC}). However, we find modulation at 14.625\,$d^{-1}$ which stands out from the local periodogram background---implying a period of 1.641 hours. This periodicity persists regardless of the choice of the detrending technique (we tried polynomial and piecewise linear), can be recovered from non-overlapping sub-sections of the TESS light curve, and is not associated with any known instrumental feature. The signal at 33.375\,$d^{-1}$ is an alias of the 1.641\,h period, given the 30\,min = 48.0\,$d^{-1}$ cadence of the light curve ($48.0\,d^{-1} - 14.625\,d^{-1} = 33.375\,d^{-1}$). This 1.641\,h period could be associated with the orbital period of the system.

An orbital period of 1.64\,hr falls below the period gap of cataclysmic variables (e.g., Ref.\cite{Knigge_etal_2011}). Another possibility is that the 1.64\,hr period represents ellipsoidal modulation of the secondary star, and the true orbital period is twice this, 3.28\,hr (which can also be identified in the periodogram at $\approx$ 7.2\,d$^{-1}$).
In either case, a period of 1.6 or 3.3\,hr, implies that the donor star must be a dwarf, consistent with the quiescent luminosity of the system. 

Ref.\cite{Worters_etal_2008} have detected magnitude variations of up to 0.32\,mag in the light curve of nova RS Oph on timescales of 600--7000\,s, 241 days after its 2006 eruption and they have suggested that this variability is due to the resumption of accretion. While the flickering observed in the TESS light curve is on a  longer timescale ($\lesssim$ 1 day), it could also be due to the resumption of accretion on the surface of the white dwarf.

\section{Optical spectroscopic observations and analysis}
\label{SI_2}

We obtained optical spectroscopic observations spanning from day 5 up to day 405 after eruption using the Ultraviolet and Visual Echelle Spectrograph (UVES) on the VLT UT2 telescope of the European Southern Observatory (ESO) in Paranal, Chile (Obs ID.\ 0100.D-0621 PI: Paolo Molaro; Obs ID.\ 2100.D-5048 PI: Paolo Molaro; Obs ID.\ 0103.D-0764, PI: Luca Izzo), the HERCULES spectrograph\cite{Hearnshaw_etal_2002} on the 1.0-m McLellan telescope at the University of Canterbury Mt. John Observatory (UCMJO, New Zealand), and the High Resolution Spectrograph (HRS) mounted on the Southern African Large Telescope (SALT; Refs.\cite{Barnes_etal_2008, Bramall_etal_2010, Bramall_etal_2012, Crause_etal_2014}). UVES covers the wavelength range 3800\,--\,10,000\,{\AA} at a resolution $R\,\approx\,59,000$, HERCULES covers the range 4000\,--\,10,000{\AA} at a resolution $R\,\approx\,41, 000$, and HRS was used in the HR mode covering 3800\,--\,9000\,{\AA} at a resolution $R\,\approx\,67,000$. 
Details of the spectroscopic observations, reduction, and analysis will be presented in Harvey et al.\ (2020, in prep). We also make use of publicly available high resolution spectroscopy from the Astronomical Ring for Access to Spectroscopy (ARAS; see Ref.\cite{Teyssier_2019}). The data, which can be found via the link \url{http://www.astrosurf.com/aras/Aras_DataBase/Novae/2018_NovaCar2018.htm}, consist of high-resolution ($R \approx$ 15,000) spectra over a limited wavelength range centred on H$\alpha$.

\subsection{Reddening.}
The Galactic reddening maps of Ref.\cite{Schlafly_Finkbeiner_2011} indicate $A_V = 3.5$ mag in the direction of nova V906~Car, which should be regarded as an upper limit for a nearby Galactic object. 

\begin{figure}
\begin{center}
  \includegraphics[width=0.47\textwidth]{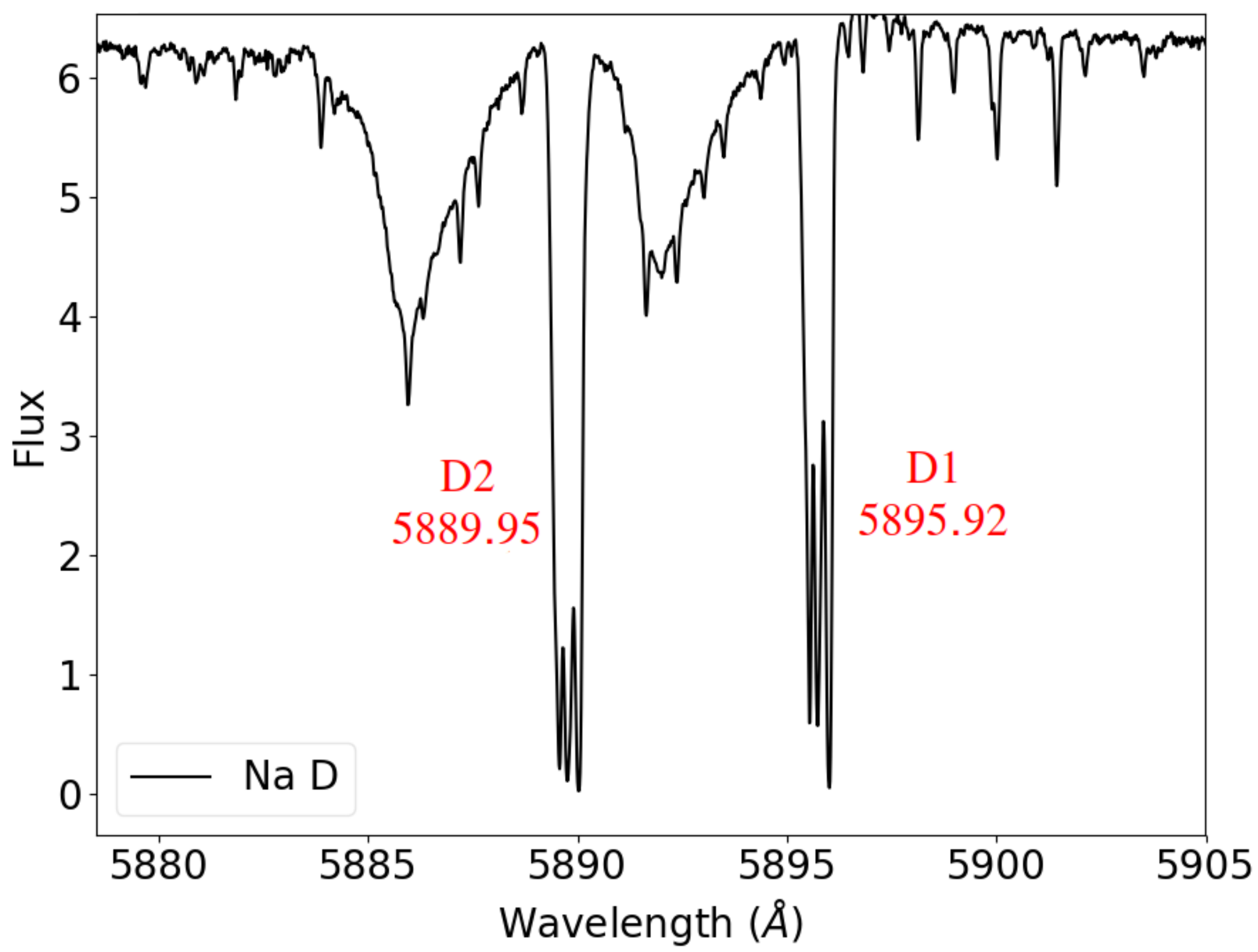}
    \includegraphics[width=0.48\textwidth]{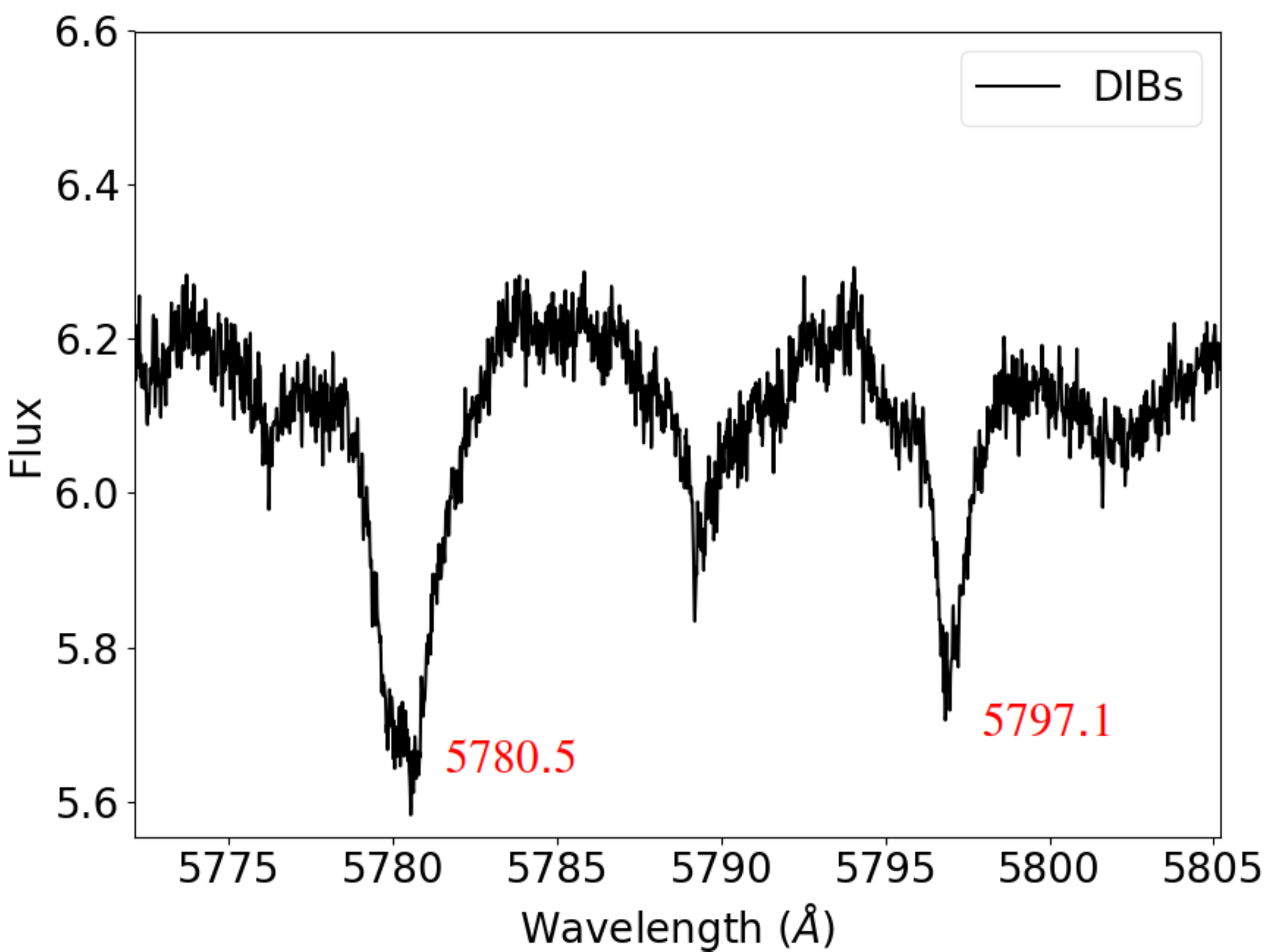}
\caption{\textbf{The spectral lines used to estimate the reddening}. \textit{Left:} the \eal{Na}{i} D interstellar absorption lines at 5895.92 \AA\ (D1) and 5889.95 \AA\ (D2). \textit{Right:} a sample of some of the diffuse interstellar bands (DIBs) used to estimate the reddening.}
\label{Fig:NaD_DIBs}
\end{center}
\end{figure}

We use the relations from Ref.\cite{Poznanski_etal_2012} to derive the extinction from the equivalent width (EW) of the \eal{Na}{i} D1 and D2 interstellar absorption doublet at 5895.92\,$\mathrm{\AA}$ and 5889.95\,$\mathrm{\AA}$, respectively (Supplementary Figure~\ref{Fig:NaD_DIBs}). We measure EW(D1) = 0.54 $\pm$ 0.02\,$\mathrm{\AA}$ and EW(D2) = 0.69 $\pm$ 0.02\,$\mathrm{\AA}$ in the spectrum taken on day 6 (see SI.\ref{SI_2}). Thus, we derive $E(B-V) = 0.38 \pm 0.05$ mag and $A_V = 1.18 \pm 0.05$ mag for $R_V = 3.1$. The reddening derived from the \eal{Na}{I} D interstellar absorption is three times smaller than the integrated total for the line of sight in the Galactic reddening maps of Ref.\cite{Schlafly_Finkbeiner_2011}, which can be explained by a moderate distance towards V906~Car.

We also use the diffuse interstellar bands (DIBs; Supplementary Figure~\ref{Fig:NaD_DIBs}) in the optical spectrum taken on day 12 with UVES to derive an estimate of the reddening from the  empirical relations of Ref.\cite{Friedman_etal_2011}.
The $E(B-V)$ values derived from the different DIBs are listed in Supplementary Table~\ref{table:DIBs}, with an average value $E(B-V) = 0.34 \pm 0.04$ mag, implying $A_V = 1.05 \pm$ 0.04. These values are in good agreement with the estimate from the \eal{Na}{i} D interstellar lines. Based on both methods, we adopt average values of $E(B-V) = 0.36 \pm 0.05$ mag and $A_V = 1.11 \pm 0.05$ mag, which we use throughout this paper. 

\begin{table}
\centering
\caption{\textbf{The EW and $E(B-V)$ derived from the DIBs.}}
\begin{tabular}{ccc}
\hline
DIB $\lambda$ & EW  & $E(B-V)$\\ 
(\AA) & (m\AA) & (mag)\\
\hline
5780.5 & 170 $\pm$ 10 & 0.33 $\pm$ 0.02\\
5797.1 & 75  $\pm$ 5  & 0.40 $\pm$ 0.06\\
6196.0 & 35  $\pm$ 5  & 0.33 $\pm$ 0.04\\
6613.6 & 60  $\pm$ 10 & 0.30 $\pm$ 0.05\\
\hline
\end{tabular}
\label{table:DIBs}
\end{table}

\begin{figure}
\begin{center}
  \includegraphics[width=0.95\textwidth]{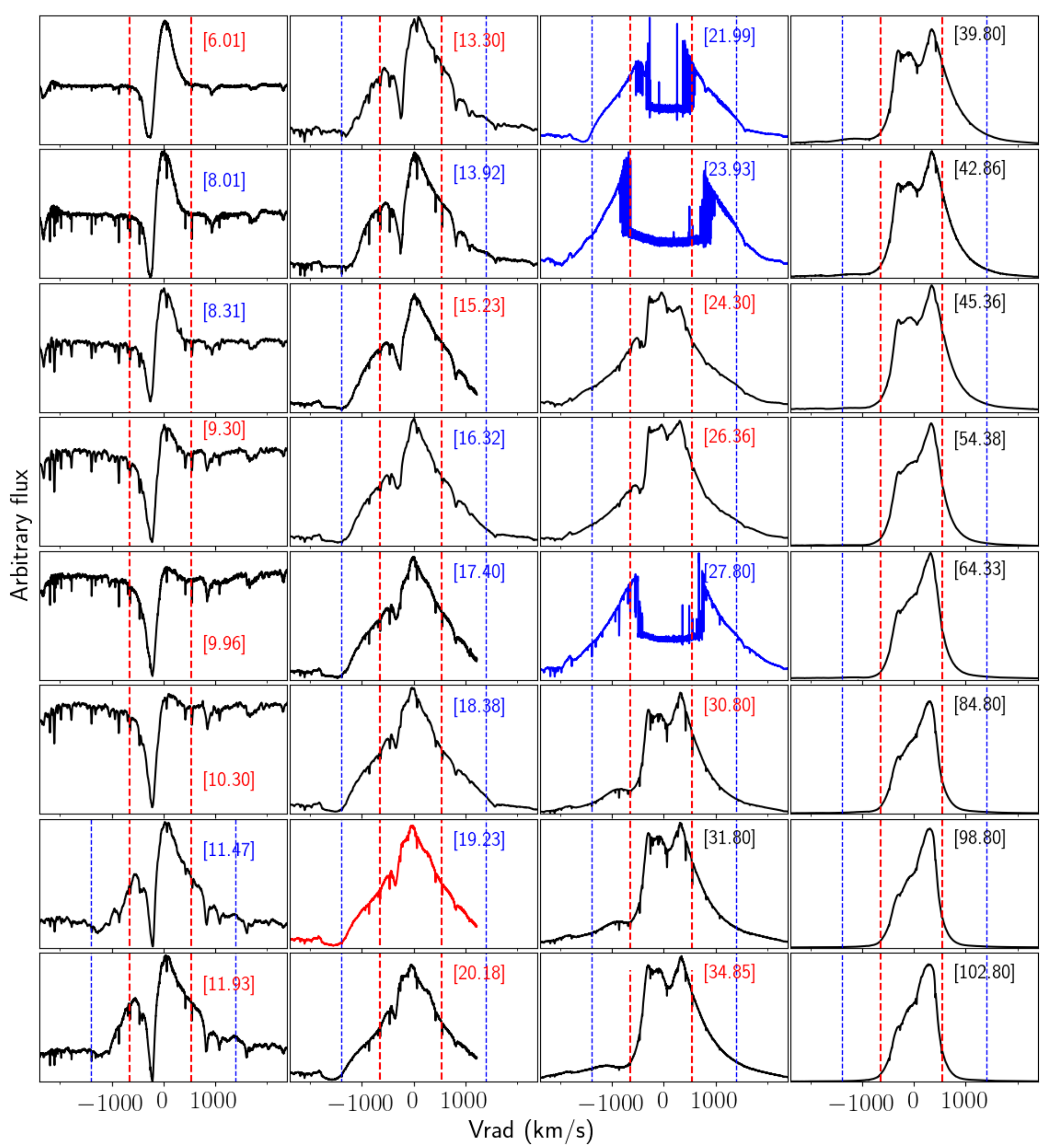}
\caption{\textbf{The evolution of the H$\alpha$ line profiles of V906~Car. The numbers between brackets are days after $t_0$; these numbers are highlighted in red if the observation coincided with a flare in the light curve. The red vertical dashed lines represent the width of the first component ($v_1$). The blue vertical dashed lines represent the width of the broad base component ($v_2$).} The profiles shown in blue with a centred ``notch'' are saturated and are shown to illustrate the change of the width of the full base component. The profile shown in red indicates the date when the light curve reversed its declining trend. Heliocentric corrections are applied to all radial velocities.}
\label{Fig:Halpha_profiles}
\end{center}
\end{figure}


\subsection{Spectral evolution.}

The spectra of V906~Car near maximum light were dominated by P Cygni profiles of \eal{H}{I}, \eal{Fe}{II}, \eal{O}{I}, and \eal{Na}{I}, along with a large number of Transient Heavy Element Absorption\cite{Williams_etal_2008,Williams_Mason_2010} (THEA) features.



\begin{figure}[!t]
\begin{center}
  \includegraphics[width=0.9\textwidth]{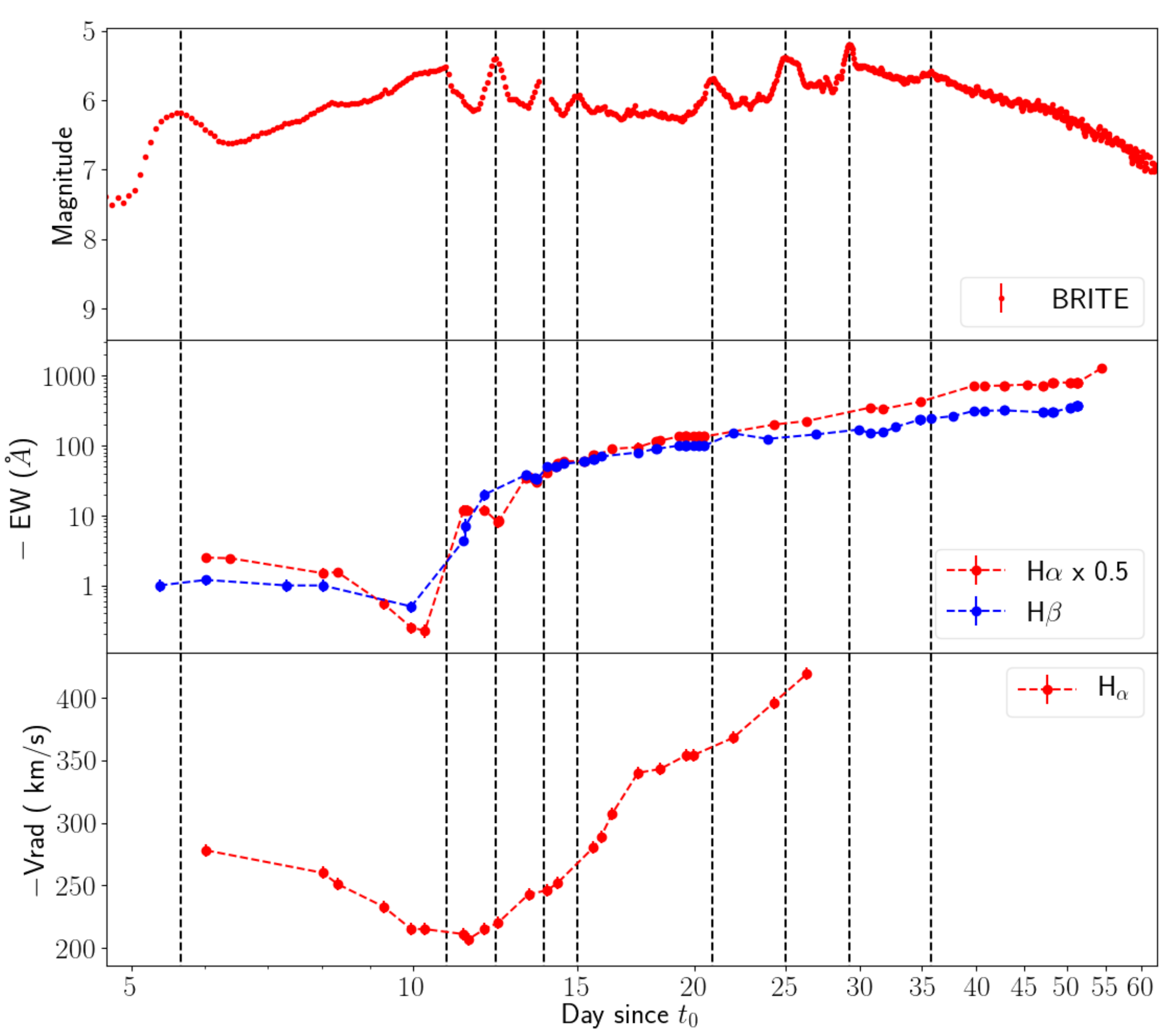}
\caption{\textbf{The evolution of the Balmer line profile EWs and velocities}. \textit{Top}: the optical light curve of V906~Car. \textit{Middle:} the evolution of the EW for H$\alpha$ and H$\beta$.  \textit{Bottom}: the evolution of the heliocentric radial velocities of the absorption component of H$\alpha$ during the flaring phase. Note that both the EWs and radial velocities are multiplied by ($-1$).}
\label{Fig:LC_EW_abs}
\end{center}
\end{figure}

Supplementary Figure~\ref{Fig:Halpha_profiles} shows the evolution of the H$\alpha$ line profiles from day 6--103. Before the first maximum, the line profile had a P Cygni form, characterized by a full-width at zero intensity, FWZI $= 1200 \pm 50$\,km\,s$^{-1}$. The minimum of the absorption component is at a heliocentric radial velocity of $-260 \pm 10$\,km\,s$^{-1}$, while the maximum expansion velocity is $v_1 \approx$ 500--600\,km\,s$^{-1}$. 
Near the first maximum, on day 10.4, the emission component weakens considerably relative to the continuum. Then, after the first maximum (day $>$ 11), the line profile shows several components: the previously existing narrow P Cygni profile on top of broad base with FWZI $= 2400 \pm 100$\,km\,s$^{-1}$. The maximum expansion velocity of the broad component is $v_2 \approx$ 1200\,km\,s$^{-1}$. On day 22, the broad base showed a significant broadening in just one day, to a FWZI of more than 5000\,km\,s$^{-1}$. At this stage, the maximum expansion velocity is $v_3 \approx$ 2500\,km\,s$^{-1}$. After day $\sim$24, the EW of the narrower component ($v_1$) increased drastically, dominating the broad base emission component. In summary, we identify at least three velocity components in the line profiles, with maximum expansion velocities of $\approx$ 600, 1200, and 2500\,km\,s$^{-1}$. These components coexist during some stages, indicating the presence of multiple, physically-distinct ejecta of different velocities.

The evolution of the EWs of H$\alpha$ and H$\beta$ are presented in Supplementary Figure~\ref{Fig:LC_EW_abs}. There is a trend of decreasing EW (increasing emission line flux compared to the continuum flux) after the first maximum. 
The evolution of the heliocentric radial velocity of the H$\alpha$ absorption line is also presented in Supplementary Figure~\ref{Fig:LC_EW_abs}. The velocity, measured from the dip of the absorption, shows an initial deceleration during the rise to the first optical maximum, which might be explained by homologously-expanding ejecta, where the origin of the absorption moves to deeper regions characterized by lower velocities. Alternatively, the initial slow component could still be gravitationally bound to the system, and therefore decelerates. The velocity reaches a minimum on day $11.5$ of around 205\,km\,s$^{-1}$, 
and then shows rapid acceleration. The acceleration might be due to the collision between the slow initial outflow with subsequent faster ejections.



\section{Insights from Radio Observations}
\label{SI_3}

We monitored V906~Car at radio wavelengths using the Australia Telescope Compact Array (ATCA), starting on 2018 April 3 (day 18). Radio monitoring continued through the time of publication, with the most recent epoch on 2019 August 31 (day 533). All observations were obtained in continuum mode with the Compact Array Broadband Backend (CABB). In the first month of outburst, observations covering 4.5--6.5 GHz (C-band) and 8--10 GHz (X-band) were obtained. Starting on 2018 October 5, lower frequency 1--3 GHz observations were also obtained. 

As of August 2019, the radio flux of V906~Car continues to increase, and the radio SED remains optically thick with a spectral index of $\alpha = 1.1 \pm 0.1$ (where $S_{\nu} \propto \nu^{\alpha}$). We model the light curve as free-free emission from an optically-thick expanding blackbody with a constant temperature of $10^4$ K. While this is the ``classic" model for radio emission from novae\cite{Seaquist_Bode_2008}, we note that it predicts a substantially steeper spectrum than observed ($\alpha = 2$). This discrepancy between the observed and expected optically-thick spectral index is by no means unique to V906~Car, but is observed in practically all novae on the radio rise. While no satisfactory explanation for the discrepancy exists, it is likely due to the density profile of the outermost nova ejecta. We also assume that the ejecta are distributed as a ``Hubble Flow" or homologous expansion (i.e., $v \propto r$). We assume a density profile $\rho \propto v^{-2}$, and must assume or fit for a minimum ejecta velocity (v$_{\rm min}$), and a maximum ejecta velocity (v$_{\rm max}$)\cite{Seaquist_Palimaka1977, Hjellming_etal_1979}.

\begin{figure}[!t]
\centering
\includegraphics[width=0.85\textwidth]{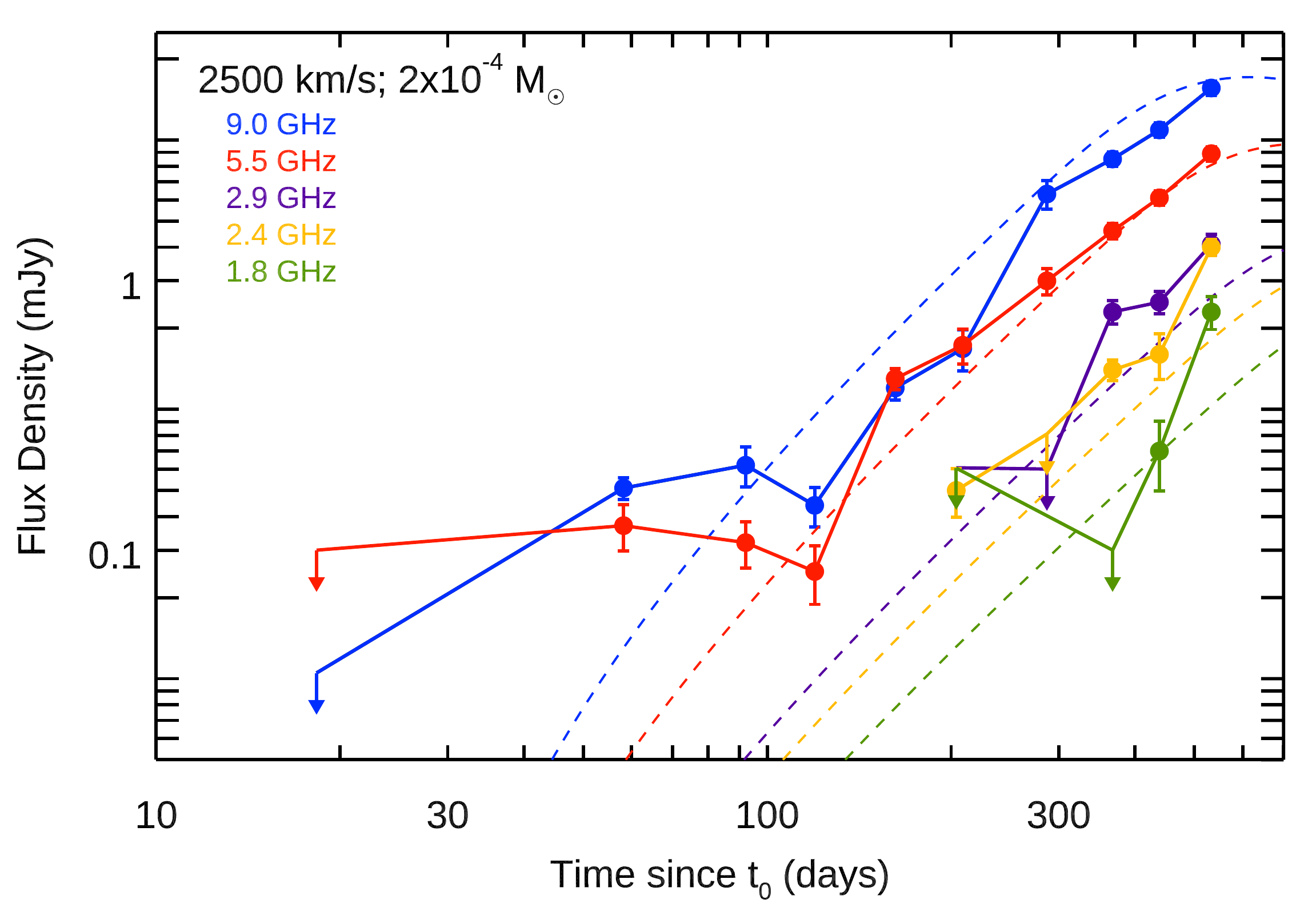}
\caption{{\bf The multi-frequency radio light curve of V906~Car over the first 533 days of outburst, superimposed with a model of thermal ejecta expanding at 2500 km s$^{-1}$.} Measurements are solid dots, or arrows marking 3$\sigma$ upper limits for non-detections; measurements at the same frequency are connected by solid lines. The model light curves at the same frequencies are plotted as dotted lines. The model represents free-free emission from isothermal  ejecta of $T_e = 10^4$ K, $M_{\rm ej} = 2.0 \times 10^{-4}$ M$_{\odot}$, $f_V = 0.006$, v$_{\rm min}$ = 500  km~s$^{-1}$, v$_{\rm max}$ = 2500 km~s$^{-1}$, at a distance of 6.2\,kpc, and assumes that expansion began on day 22 of the outburst. The error bars are 1$\sigma$ uncertainties.}
\label{Fig:radiolc_2500}
\end{figure}

Radio free-free luminosity scales with emission measure, and so depends on the clumping of the ejecta or the ``filling factor".
We use the UVES spectrum obtained on day 405.0 after the eruption (2019 April 25; well into the nebular phase) and the method described in Ref.\cite{Finzell_etal_2018} to derive an estimate for the filling factor. 
We obtain a volume filling factor of $f_{V} = 0.49 \pm 0.07$ assuming an ejecta velocity of $v_{\mathrm{ej}}$ = 600 km\,s$^{-1}$, $f_{V} = 0.06 \pm 0.01$ assuming $v_{\mathrm{ej}}$ = 1200 km\,s$^{-1}$, and $f_{V} = 0.006 \pm 0.001$ assuming $v_{\mathrm{ej}}$= 2500 km\,s$^{-1}$. A detailed description of the calculation of the filling factor will be presented in Harvey et al.\ (2020, in prep.). See also Ref.\cite{Molaro_etal_2019} for further spectroscopic analysis.

If we assume that the radio-emitting ejecta are expanding with v$_{\rm max}$ = 2500 km s$^{-1}$ starting on day 22, we find that V906~Car must either be at the far end of our estimated distance range ($\sim$6.2 kpc) or its ejecta must be substantially cooler than $10^4$ K to not over-predict the radio fluxes. The model light curve is superimposed on our radio observations in Supplementary Figure \ref{Fig:radiolc_2500}. In order for the light curve to remain optically thick for at least 533 days at frequencies as high as 9 GHz, and assuming a filling factor of $f_V = 0.006$, we find that the ejecta mass must be $M_{\rm ej} \gtrsim 2 \times 10^{-4}$ M$_{\odot}$. Continued monitoring of the light curve as it peaks and turns over will enable us to move from an upper limit on the mass to a (model dependent) mass measurement.

If instead the radio-emitting ejecta are expanding with v$_{\rm max}$ = 1200 km s$^{-1}$ starting on day 11, the radio light curve implies a distance to V906~Car of 3.4\,kpc (assuming $T_e = 10^4$ K). Taking a filling factor $f_V = 0.06$, the long duration of the optically thick rise implies an ejecta mass $>1.5 \times 10^{-4}$ M$_{\odot}$. Similarly, if the radio emission originates with the slowest ejecta (v$_{\rm max}$ = 600 km s$^{-1}$), the implied distance is 1.8\,kpc and the ejecta mass is $>9 \times 10^{-5}$ M$_{\odot}$.
These lower values of v$_{\rm max}$ yield very similar model radio light curves to the ones shown in Supplementary Figure \ref{Fig:radiolc_2500}, and the larger filling factors derived for lower v$_{\rm max}$ imply similar upper limits on the ejecta mass, independent of v$_{\rm max}$.

In either scenario, there is a clear indication of excess radio emission between days 50 and 100, above the expectation for the optically-thick expanding blackbody. Similar early time excesses have been seen in several other novae---both detected by \emph{Fermi}/LAT\cite{Chomiuk_etal_2014, Finzell_etal_2018} and non-detected\cite{Taylor_etal_1987, Krauss_etal_2011, Weston_etal_2016_v1723}. The leading explanation for this excess is synchrotron emission produced by relativistic electrons accelerated in nova shocks \cite{Chomiuk_etal_2014, Vlasov_etal_2016}. The first radio detection occurred on day 58, $\sim$10 days after the last \emph{Fermi}/LAT detection. It is likely that the radio synchrotron emission originated with the same shock as the $\gamma$-rays, but that the ejecta were optically thick at radio wavelengths during our first ATCA observation (day 18), yielding a non-detection. By day 58, the optical light curve has smoothed out and is squarely on its decline (Supplementary Figure \ref{Fig:sub}), implying thinning ejecta that would enable the escape of synchrotron emission from the internal shock.

\section{Optical--$\gamma$-ray correlation.}
\label{SI_4}

The exceptional $\gamma$-ray brightness of V906~Car allowed us to construct the best sampled $\gamma$-ray light curve for a nova to date.
The $\gamma$-ray light curve shows a series of flares (on days 25, 29, and 36) coinciding with the peaks in the optical light curve. We characterize the $\gamma$-ray flares in the bottom portion of Supplementary Table~\ref{table:flares}. The time of each flare's maximum, in units of days after $t_0$, is $t_{{\rm peak},\gamma}$ and has an associated uncertainty of $\pm0.25$ day. Meanwhile, $\Delta t_{\gamma}$ is the duration between the start of the rise and the time when the brightness drops to the same level as before the rise, with an uncertainty of $\pm0.5$ days. $F_{{\rm peak},\gamma}$ is the $\gamma$-ray flux at the flare peak, and $\Delta F_{\gamma}$ measures the change in flux between the flare peak and inter-flare levels.


To measure the possible time lag between the optical and $\gamma$-ray flares we use two techniques: the discrete cross-correlation function and the smoothness parameter.
The discrete cross-correlation function\cite{1988ApJ...333..646E} has been the standard tool for this kind of analysis for decades. If there is a time lag between the two light curves, shifting one of the light curves in time by this lag should result in the highest value of the correlation coefficient between the measurements taken close in time at the two bands. The other technique relies on the idea that the composite light curve---combining measurements in the two bands---should be the smoothest if the time lag between the light curves has been properly accounted for. The measure of smoothness for cross-correlation analysis can take into account the different scaling of the two light curves\cite{2017ApJ...844..146C} by using the ratio of the mean-square difference between consecutive light curve points to the light curve variance\cite{vonneumannetal1941,vonneumann1941,vonneumann1942}.
According to simulations\cite{2017ApJ...844..146C}, the smoothness parameter outperforms the discrete cross-correlation function, resulting in more accurate estimates of the time lag.

\begin{figure}
\begin{center}
  \includegraphics[width=0.48\textwidth]{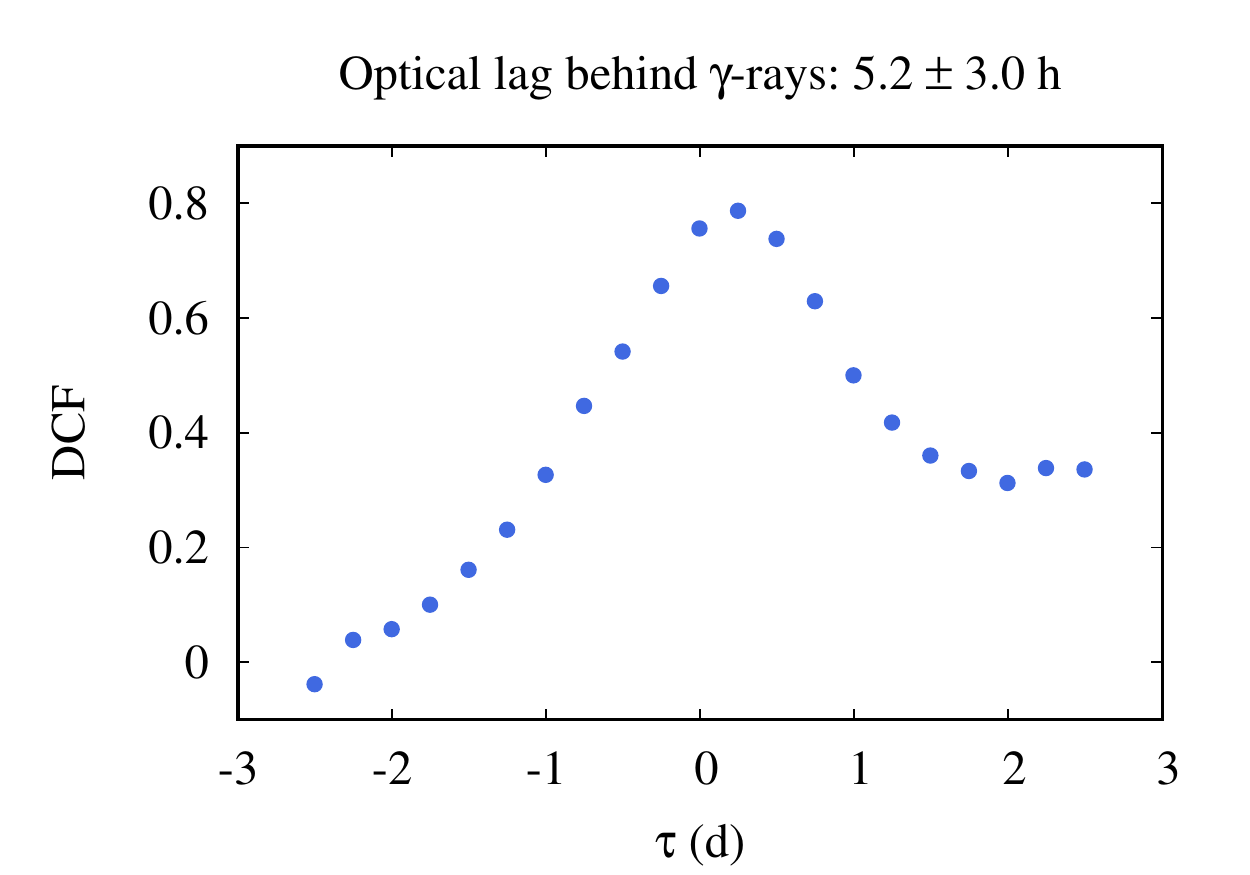}
  \includegraphics[width=0.48\textwidth]{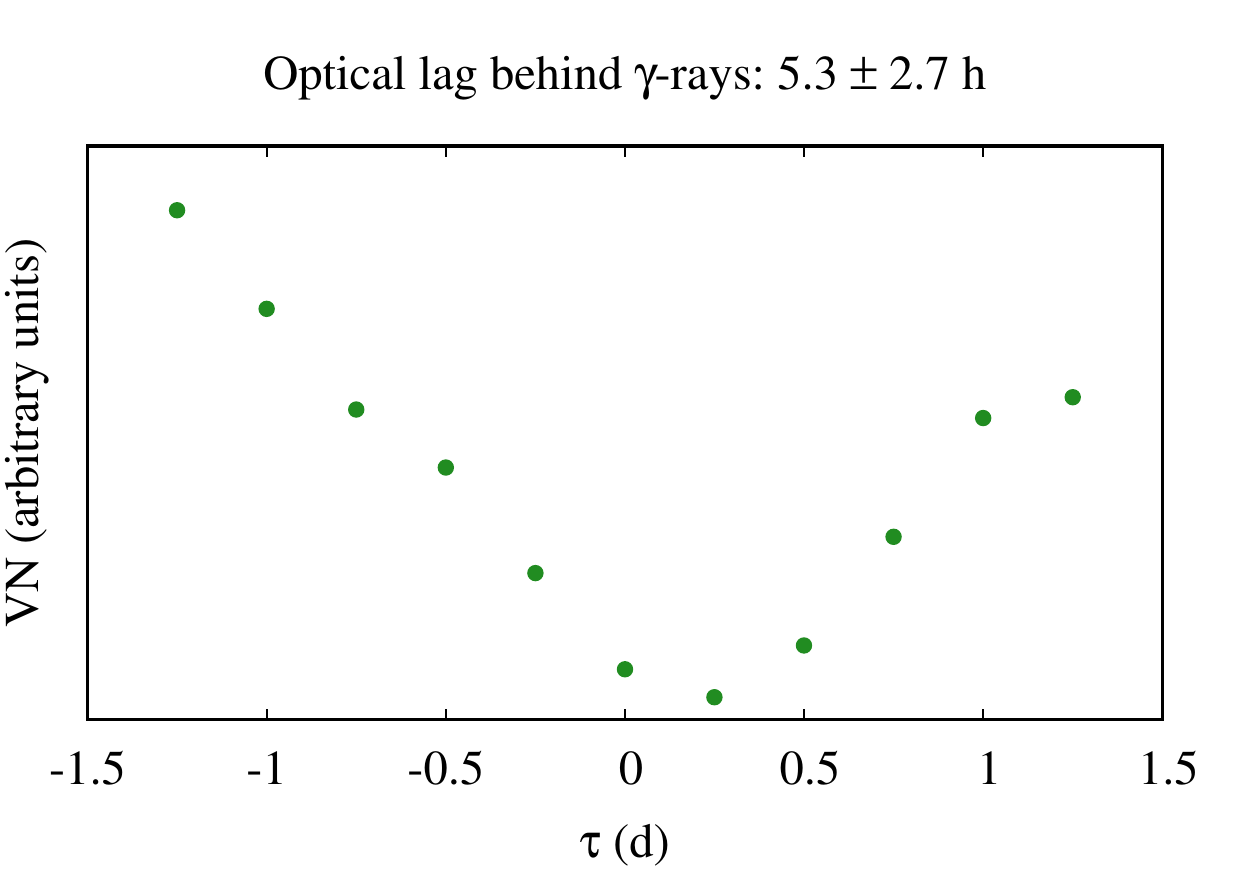}  
\caption{\textbf{Cross-correlation between the BRITE optical and the \textit{Fermi}-LAT $\gamma$-ray light curves} characterized by the binned discrete cross-correlation function (DCF\cite{1988ApJ...333..646E}; left panel) and the smoothness parameter (von Neumann estimator\cite{2017ApJ...844..146C}; right panel). The peak of the DCF and the minimum of the smoothness parameter VN correspond to the most significant correlation.}
\label{Fig:croscorr}
\end{center}
\end{figure}

When applied to the BRITE and \textit{Fermi}-LAT light curves of V906~Car, the two techniques provide consistent results. The $\gamma$-ray emission leads the optical emission by $5.3 \pm 2.7$\,hr based on the smoothness analysis, and $5.2 \pm 3.0$\,hr using the discrete cross-correlation function at 2-$\sigma$. 
The uncertainties are estimated by generating $10^4$ pairs of optical/$\gamma$-ray light curves with each real flux measurement replaced by random draws from a Gaussian distribution where the mean equals the actual measured flux and the variance equals the estimated uncertainty of the flux measurement squared. The quoted uncertainties are the standard deviations of the time lag measurements resulting from the cross-correlation analysis of these simulated pairs of light curves.

\begin{figure}[!t]
\begin{center}
  \includegraphics[width=0.8\columnwidth]{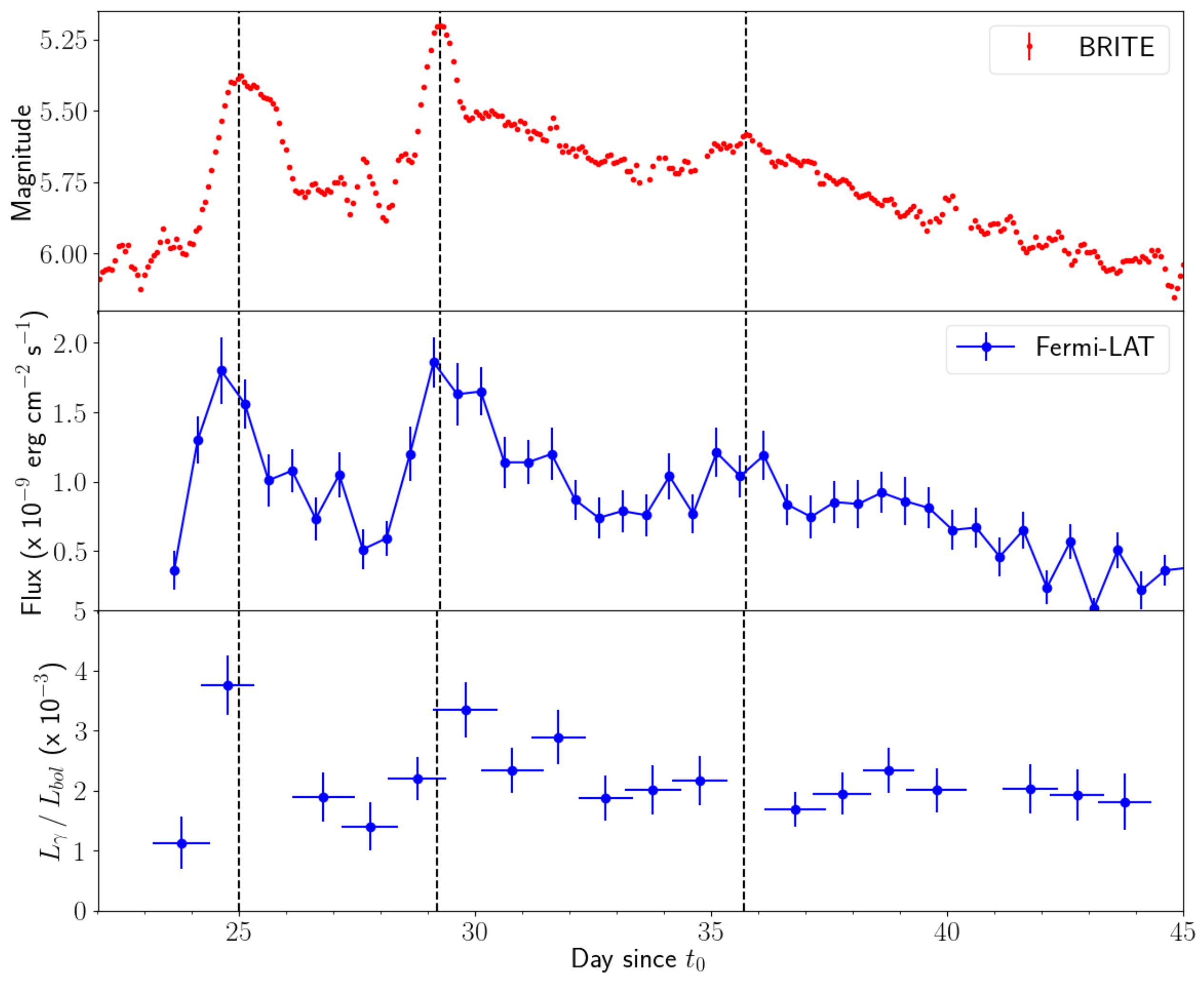}
\caption{\textbf{The evolution of $L_{\gamma}/L_{\mathrm{bol}}$ (\textit{bottom)} in comparison with the BRITE optical light curve (\textit{top}) and the \textit{Fermi}-LAT $\gamma$-ray light curves (\textit{middle}).} The black dashed lines represent the dates of the flares. The reported error bars are 1$\sigma$ uncertainties.}
\label{Fig:Lum_ratio}
\end{center}
\end{figure}

\section{Comparison of $\gamma$-ray Luminosity with  optical/X-ray Luminosities}
\label{SI_5}

\subsection{Bolometric to $\gamma$-ray luminosity ratio.}
An important diagnostic of the $\gamma$-ray producing shocks is the fraction of the nova's bolometric luminosity emerging in the $\gamma$-ray band\cite{Metzger_etal_2015, Li_etal_2017_nature}. In the first $\sim$40 days of V906~Car's outburst, the optical light curve plateaued near its maximum light value, and the SED  peaks in the optical band throughout this time (Supplementary Figure \ref{Fig:SEDs}).
Therefore, to estimate the bolometric luminosity of the nova outburst as a function of time, we use the 
temperatures derived from the SED fitting (Supplementary Figure \ref{Fig:colours}), 
Ccombined with the bolometric corrections tabulated in Ref.\cite{Weidemann_1967}. The bolometric corrections in Ref.\cite{Weidemann_1967} assume a blackbody emission, which is not an ideal assumption and should be used with caution (Ref.\cite{Gallagher_Starrfield_1976}), particularly since the spectra of novae can be dominated by emission lines. However, the SEDs of nova V906~Car are well described by a blackbody near the optical peak, therefore we use this assumption to derive the bolometric luminosity, while noting the above caveats.

The evolution of the ratio between the bolometric and $\gamma$-ray luminosity is presented in Supplementary Figure~\ref{Fig:Lum_ratio}. We find that the $\gamma$-ray luminosity is 0.2--0.4\% of the bolometric luminosity during day 24--44. These values are comparable to other novae with $\gamma$-ray emission (see Supplementary Figure~\ref{Fig:Lum_ratio_comp}). $L_{\gamma}/L_{\mathrm{bol}}$ shows a slight increase during the flares, which suggests that 
the shocks are only powering a fraction of the bolometric luminosity, as discussed below. 

The remarkable correlation between the $\gamma$-ray luminosity and the optical luminosity can be explained by a simple model in which the optical luminosity is the sum of a constant luminosity source and luminosity that is created by the time-dependent shocks internal to the ejecta. Assuming that a constant fraction ($\varepsilon$) of the shock-powered bolometric luminosity goes into $\gamma$-ray luminosity, the ratio between the $\gamma$-ray luminosity and the bolometric luminosity can be fit with a simple linear relation, $L_\gamma=\varepsilon(L_{\rm bol}-L_{\rm const})$. We find that the best fit gives a $\gamma$-ray efficiency of $\varepsilon = (4.1\pm 0.7)\times10^{-3}$, and a constant luminosity level of $L_{\rm const}=(1.6\pm0.6)\times 10^{38}\, (d/4.0\,{\rm kpc})^2$ erg s$^{-1}$ (Supplementary Figure~\ref{Fig:linear_fit}). The most likely source of $L_{\rm const}$ is residual nuclear burning on the surface of the white dwarf, which yields similar luminosities to our estimate of $L_{\rm const}$\cite{Paczynski_1971,Wolf_etal_2013}.

\begin{figure}[!t]
\begin{center}
  \includegraphics[width=0.9\columnwidth]{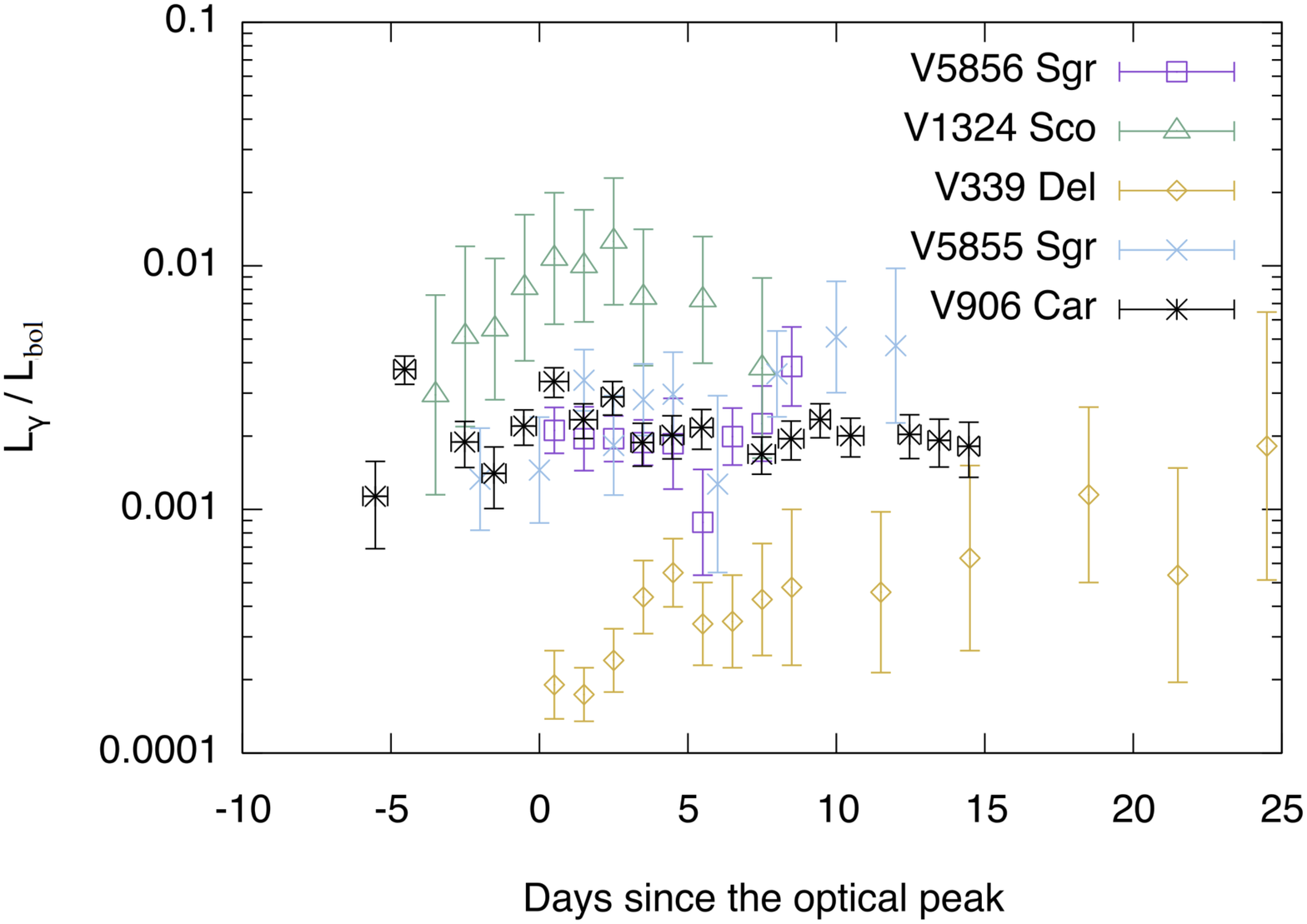}
\caption{\textbf{The evolution of $L_{\gamma}/L_{\mathrm{bol}}$ of V906 Car in comparison with other $\gamma$-ray emitting novae.} 
The ratios of V5856 Sgr (ASASSN-16ma) and V5855 Sgr were taken from Ref.\cite{Li_etal_2017_nature}, while the ratios of V1324 Sco and V339 Del were taken from Ref.\cite{Metzger_etal_2015}. The time zeros are set at the optical peaks, which are on June 20, 2012 (V1324 Sco \cite{Cheung_etal_2016}), August 16, 2013 (V339 Del \cite{Cheung_etal_2016}), October 31, 2016 (V5855 Sgr \cite{Li_etal_2017_nature}), and November 8, 2016 (V5856 Sgr \cite{Li_etal_2017_nature}). The reported errors are 1$\sigma$ uncertainties.}
\label{Fig:Lum_ratio_comp}
\end{center}
\end{figure}

\begin{figure}
\begin{center}
  \includegraphics[width=0.48\textwidth]{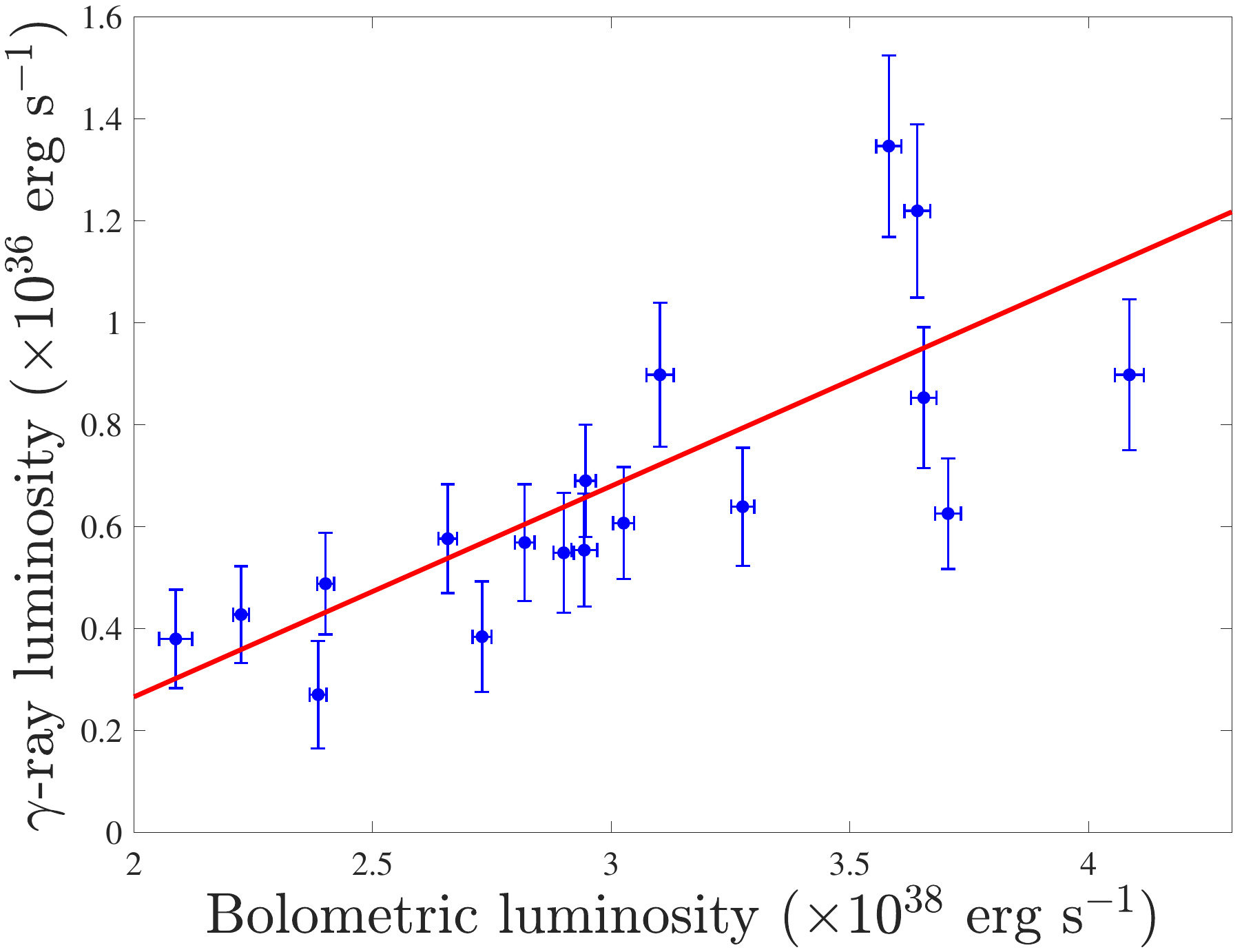}
   \includegraphics[width=0.48\textwidth]{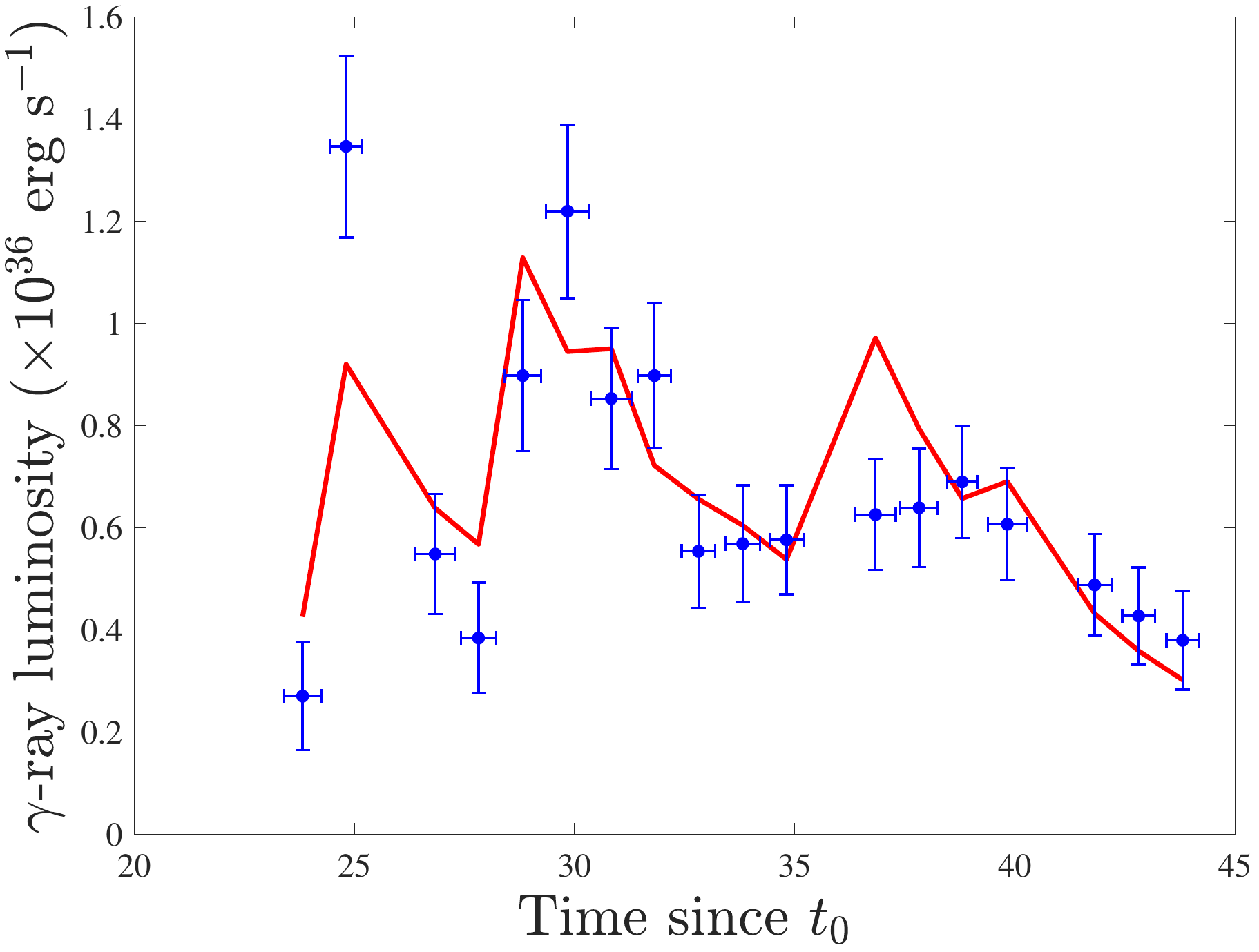}
\caption{\textit{Left:} \textbf{The $\gamma$-ray luminosity as a function of the bolometric luminosity.} Overplotted is a linear fit with the best fit parameters of $\varepsilon = (4.1\pm 0.7)\times10^{-3}$ and $L_{\rm WD}=(1.3\pm0.6)\times 10^{38}$ erg~s$^{-1}$.
\textit{Right:} \textbf{The derived $\gamma$-ray luminosity from our best fit linear model (red line) compared with the measured $\gamma$-ray luminosities (blue points).} The reported errors are 1$\sigma$ uncertainties.}
\label{Fig:linear_fit}
\end{center}
\end{figure}

\subsection{X-ray to $\gamma$-ray luminosity ratio.}

The presence of hard X-ray  emission in novae is often explained in terms of plasma heating by shocks\cite{2014ASPC..490..327M}. We interpret the {\em NuSTAR} detection of V906~Car as evidence for the presence of shocks deeply embedded in the nova ejecta. The X-ray/$\gamma$-ray flux ratio on day 36 is $L_{X}/L_{\gamma} = 0.013 \pm 0.003$.
If the same shock is responsible for heating the plasma observed by {\em NuSTAR} and accelerating the particles emitting the GeV $\gamma$-rays, the flux ratio on day 36 is smaller than the theoretical values predicted by the radiative shock models. These models predict that most of the shock energy should be emitted in X-rays and only a small fraction is spent on particle acceleration\cite{Metzger_etal_2015}. Possible explanations for the low $L_{X}/L_{\gamma}$ in V906~Car include (1) a corrugated shock front which produces softer X-rays which are more easily absorbed by the intervening ejecta\cite{Steinberg_Metzger_2018}, (2) lower-density than expected\cite{Metzger_etal_2014}, adiabatic (rather than radiative) shocks, (3) different shock systems responsible for the $\gamma$-ray and hard X-ray emission of the nova\cite{Nelson_etal_2019}, or (4) an (unknown) exceptionally efficient particle acceleration mechanism capable of passing {\em most} of the shock energy to non-thermal particles.

\section{Theoretical interpretation}
\label{SI_6}

\subsection{The optical-to-$\gamma$-ray lag interpretation.}
Here we discuss theoretical expectations for the relative time lag between the optical and $\gamma$-ray light curves, assuming that both are originating from shocks deeply embedded in the nova ejecta.
For the purposes of this estimate, we treat the nova ejecta as a spherical steady outflow of velocity $v_{\rm w}$, mass loss rate $\dot{M}$, and radial density profile $\rho_{\rm w} = \dot{M}/(4\pi r^{2} v_w)$. Due to the extremely short mean free path of UV or X-ray photons, the reprocessed optical radiation will originate close to the shock surface\cite{Metzger_etal_2015}. Likewise, due to the short mean free path for cosmic rays to pion produce on ambient ions in hadronic scenarios (or for electrons to interact with the nova optical light in leptonic scenarios), the $\gamma$-rays can also be effectively generated near the instantaneous position of the shock\cite{Metzger_etal_2015}.

\begin{figure}[!t]
\includegraphics[width=0.49\textwidth]{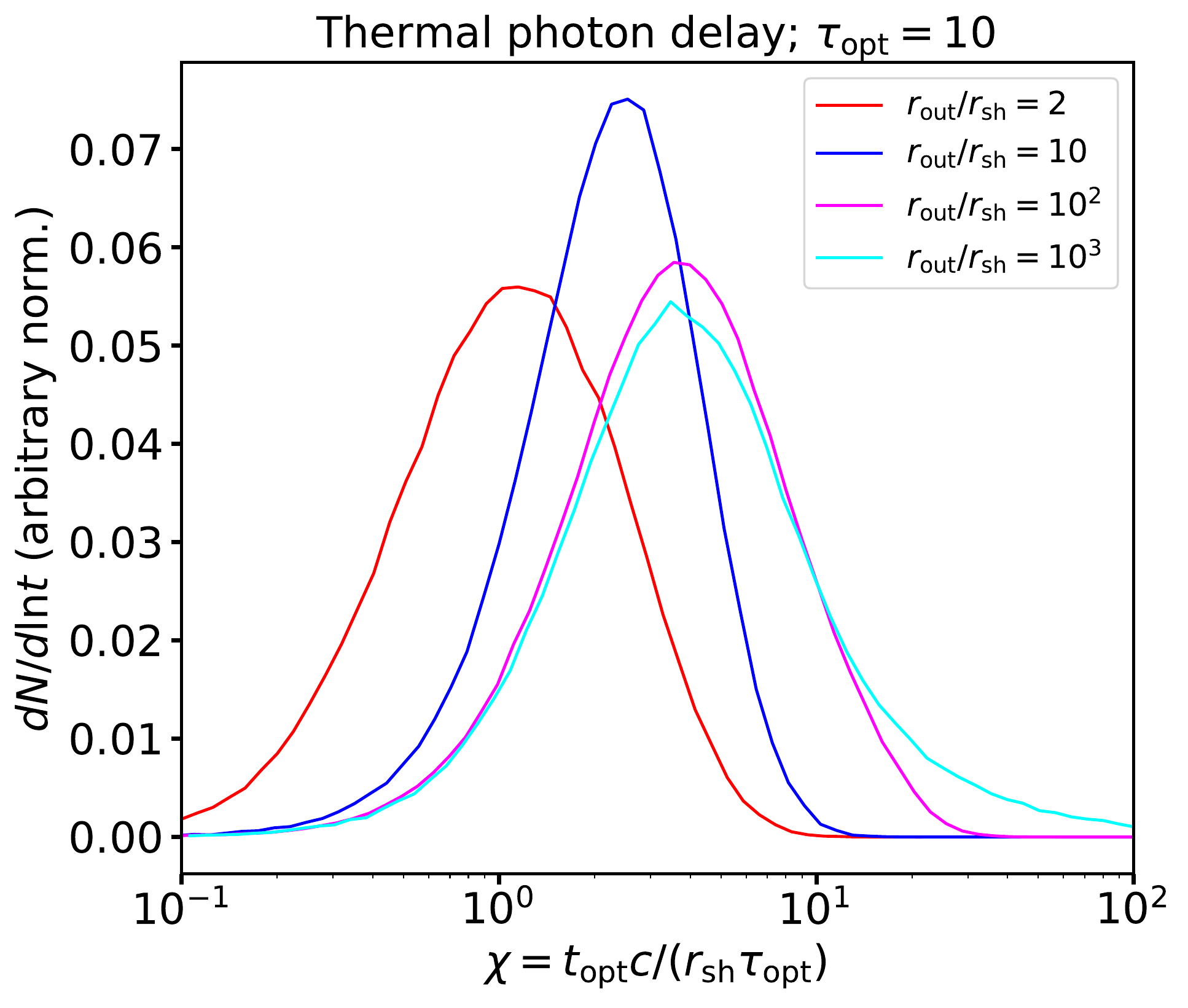}
\includegraphics[width=0.49\textwidth]{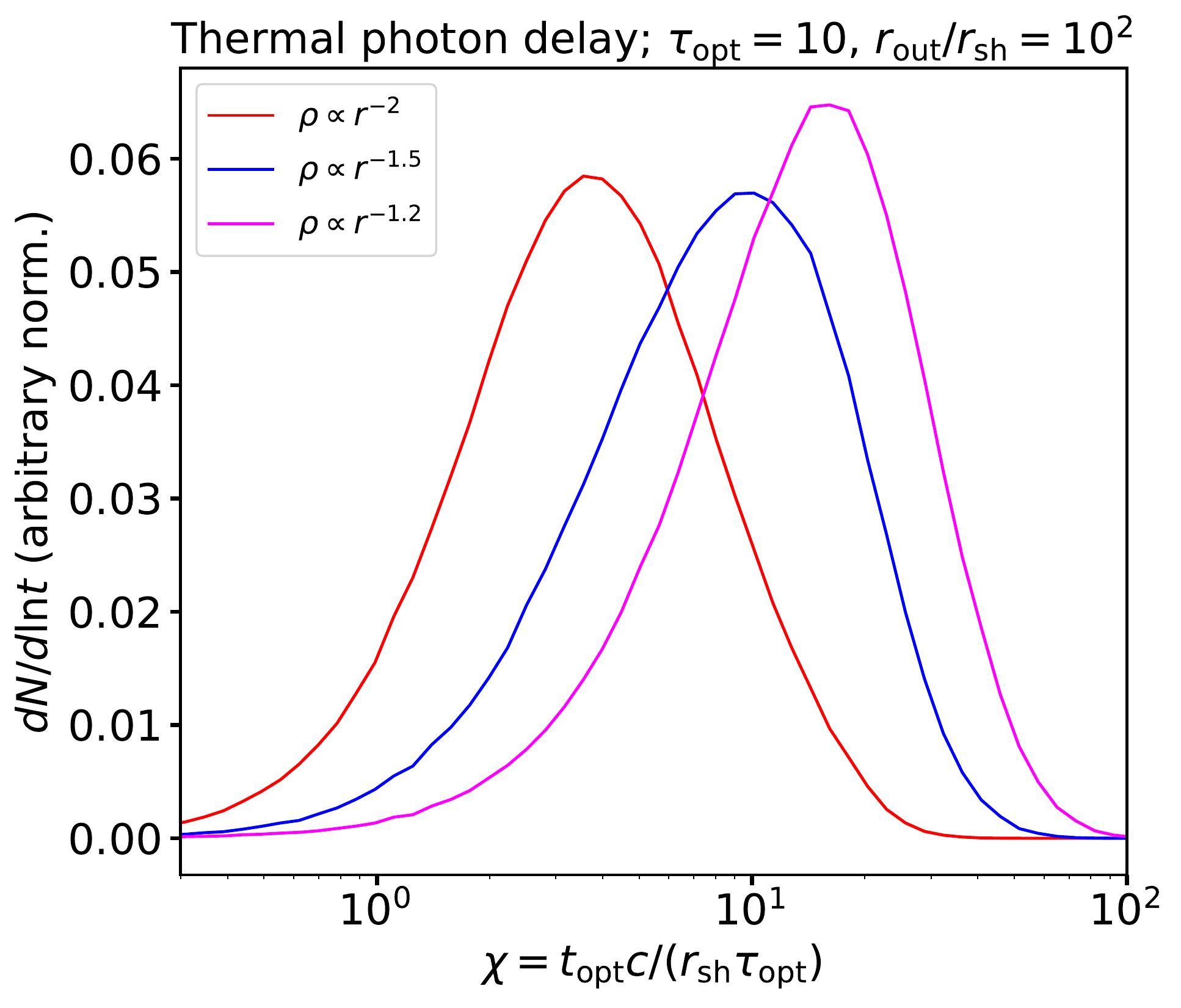}
\caption{\textbf{Distribution of optical photons according to their delay times, emitted at a deeply embedded shock at radius $r_{\rm sh}$,
after passing through the external shell of nova ejecta ahead of the shock, as calculated by means of a Monte Carlo simulation.} The delay time shown has been normalized to the characteristic photon diffusion timescale {\it at the emission site}, $\tau_{\rm opt}(r_{\rm sh}) r_{\rm sh}/c$, where $\tau_{\rm opt}$ is the total optical depth through the ejecta shell (from $r_{\rm sh}$ to $r_{\rm out}$); in other words, we are plotting a numerically-determined value of $\chi$ shown in Equations (\ref{eq:topt}, \ref{eq:chi}).  The left panel shows a case in which the external medium ahead of the shock is that of a steady wind with a radial density profile $\rho \propto r^{-2}$ extending from $r = r_{\rm sh}$ to $r_{\rm out}$. The right panel shows how $\chi$ increases for density profiles more shallow than $\rho \propto r^{-2}$, as would be expected if the mass-loss rate of the nova outflow were decreasing with time. A larger value of $\chi$ in the latter case could contribute to the $\sim 5$ hour delay in the optical relative to the $\gamma$-ray variations in V906 Car.}
\label{fig:diff}
\end{figure}

When the shocks are deeply embedded, $\gamma$-rays in the 0.1--10 GeV energy range  observed by {\it Fermi}/LAT are attenuated by electron/positron pair creation on nuclei, for which the opacity is roughly $\kappa_{\gamma} \approx 0.005-0.01$ cm$^{2}$/g, with only a weak dependence on the $\gamma$-ray energy (e.g., Ref.\cite{Zdziarski_Svensson_1989}). For a shock radius of $r_{\rm sh}$, the optical depth from the shock to the ejecta surface is $\tau_{\gamma} \approx \int_{r_{\rm sh}}^{\infty}\rho_{\gamma}\kappa_{\gamma}dr \approx \dot{M}\kappa_{\gamma}/(4\pi v_{w} r_{\rm sh})$. The $\gamma$-rays become detectable once the shock reaches the critical radius $r_{\rm sh} = r_{\gamma} = \dot{M}\kappa_{\gamma}/(4\pi v_w)$ at which $\tau_{\gamma} \lesssim 1$. This occurs on a timescale of
\begin{equation}
t_{\gamma} = \frac{r_{\gamma}}{v_{\rm sh}} \approx \frac{\dot{M}\kappa_{\gamma}}{4\pi v_{\rm sh}v_w},
\end{equation}
after the shock was launched, where $v_{\rm sh}$ is the radial velocity of the shocked gas.

Unlike $\gamma$-rays, which vanish after being absorbed, thermal optical radiation can diffuse out of the ejecta even when its optical depth is high. The timescale for optical photons to escape the ejecta is given by the photon diffusion timescale
\begin{eqnarray}
t_{\rm opt} &\approx& \chi\tau_{\rm opt}\frac{r_{\rm sh}}{c} \approx \frac{\chi\dot{M}\kappa_{\rm opt}}{4\pi c v_w} \nonumber \\
&\approx& 1\,{\rm hr}\left(\frac{\chi}{2}\right) \left(\frac{\dot{M}}{10^{-4}M_{\odot}{\rm week^{-1}}}\right)\left(\frac{v_w}{500{\rm \,km/s}}\right)^{-1}\left(\frac{\kappa_{\rm opt}}{0.1\,\,{\rm cm^{2}/g}}\right).
\label{eq:topt}
\end{eqnarray}
Here, $\tau_{\rm opt} \approx \int_{r_{\rm sh}}^{\infty}\rho_{\rm w}\kappa_{\rm opt}dr$ is the visual wavelength optical depth and $\kappa_{\rm opt}$ is the effective optical opacity (e.g., from Doppler-broadened absorption lines).
The numerical factor $\chi$ depends on the radial profile and extent of the wind. We have estimated it using Monte Carlo simulations in which we injected photons at a fixed time at the instantaneous shock radius $r_{\rm sh}$ and collected them as they escaped through $r_{\rm out}$; the distribution of photons according to their residence time between $r_{\rm sh}$ and $r_{\rm out}$ is shown in Supplementary Figure~\ref{fig:diff} for different wind parameters. For a wind with a steady mass-loss rate and velocity ($\rho_{\rm w} \propto r^{-2}$) that has lasted a time $t_w$, the optical diffusion time depends rather weakly on the extent of the wind ($r_{\rm out} = v_w\,t_w$); we find $\chi \approx 1-5$ for $r_{\rm out}/r_{\rm sh} \sim 2-10^{3}$ (Supplementary Figure~\ref{fig:diff} left panel). On the other hand, a shallower density profile, corresponding to a decreasing mass-loss rate or increasing wind velocity, further increases $\chi$ and delays the emergence of the optical photons (Supplementary Figure~\ref{fig:diff} right panel). In this case, the nominal diffusion time $\tau_{\rm opt}(r) r/c$ (i.e. the time for photons to diffuse from $r$ to $2r$) increases with $r$ and the photons spend most of their time just below the optical photosphere $r_{\rm opt}$, beyond which the diffusion approximation breaks down and the photons can free-stream out. 

The time lag between $\gamma$-ray and optical peaks from an emerging shock is
approximately given by
\begin{equation}
t_{\rm delay} \equiv t_{\rm opt}-t_{\gamma} = \frac{\dot{M}}{4\pi c v_w}\left(\chi\kappa_{\rm opt} - \kappa_{\gamma}\frac{c}{v_{\rm sh}}\right)\,. \end{equation}
Since we typically expect $\kappa_{\rm opt} \lesssim 0.1$ cm$^{2}$/g and $c/v_{\rm sh} \lesssim 100$ for characteristic shock velocities ($v_{\rm sh} \lesssim 3000$ km s$^{-1}$), we expect that $t_{\rm delay} < 0$ (i.e., the $\gamma$-ray peak should lag the optical).  However, the characteristic timescale of this delay, $\sim t_{\rm opt}$, is expected to be small, hours or less (Equation \ref{eq:topt}). This conclusion can be reversed, with the optical lagging the $\gamma$-rays, for sufficiently large values of
\begin{equation}
\chi \gtrsim \left(\frac{\kappa_{\gamma}}{\kappa_{\rm opt}}\right)\left(\frac{c}{v_{\rm sh}}\right) \approx 15 \left(\frac{v_{\rm sh}}{10^{3}\,{\rm km/s}}\right)\,.
\label{eq:chi}
\end{equation}
As shown in the right hand panel of Supplementary Figure~\ref{fig:diff}, such a large value of $\chi$ is possible for an external medium with a density profile which decays more shallowly with radius than that of a steady-wind (e.g.~$\rho \propto r^{-1.5}$), as would be expected in cases where the wind mass-loss rate is decreasing as a function of time since the beginning of the outburst. A delay time $t_{\mathrm{delay}}$ of the order of a few hours between the $\gamma$-ray and optical emission is consistent with the lag we measure from our correlation analysis in SI.\ref{SI_4}, confirming the above assumption that both $\gamma$-ray and optical emission are originating from deeply embedded shocks. 

\subsection{Leptonic vs.\ hadronic origin for the $\gamma$-rays.}

Both leptonic and hadronic processes can, in principle, contribute to the observed $\gamma$-ray emission in novae. If leptonic processes dominate, the $\gamma$-rays arise from electrons directly accelerated at the shock and emitting relativistic bremsstrahlung and inverse Compton (IC) radiation by interacting with background (thermal) matter and the optical radiation field, respectively. In hadronic models, shock-accelerated protons undergo inelastic scatterings on background hadrons and produce neutral and charged pions. Upon decay, the pions give rise to both high-energy photons as well as secondary electron-positron pairs, which contribute to the $\gamma$-ray budget via the same processes as in the leptonic scenario. Determining whether the leptonic or hadronic mechanism dominates the $\gamma$-ray emission in novae would make a substantial contribution to our understanding of the physics of particle acceleration in non-relativistic radiative shocks.

\begin{figure*}
\begin{center}
  \includegraphics[width=0.49\textwidth]{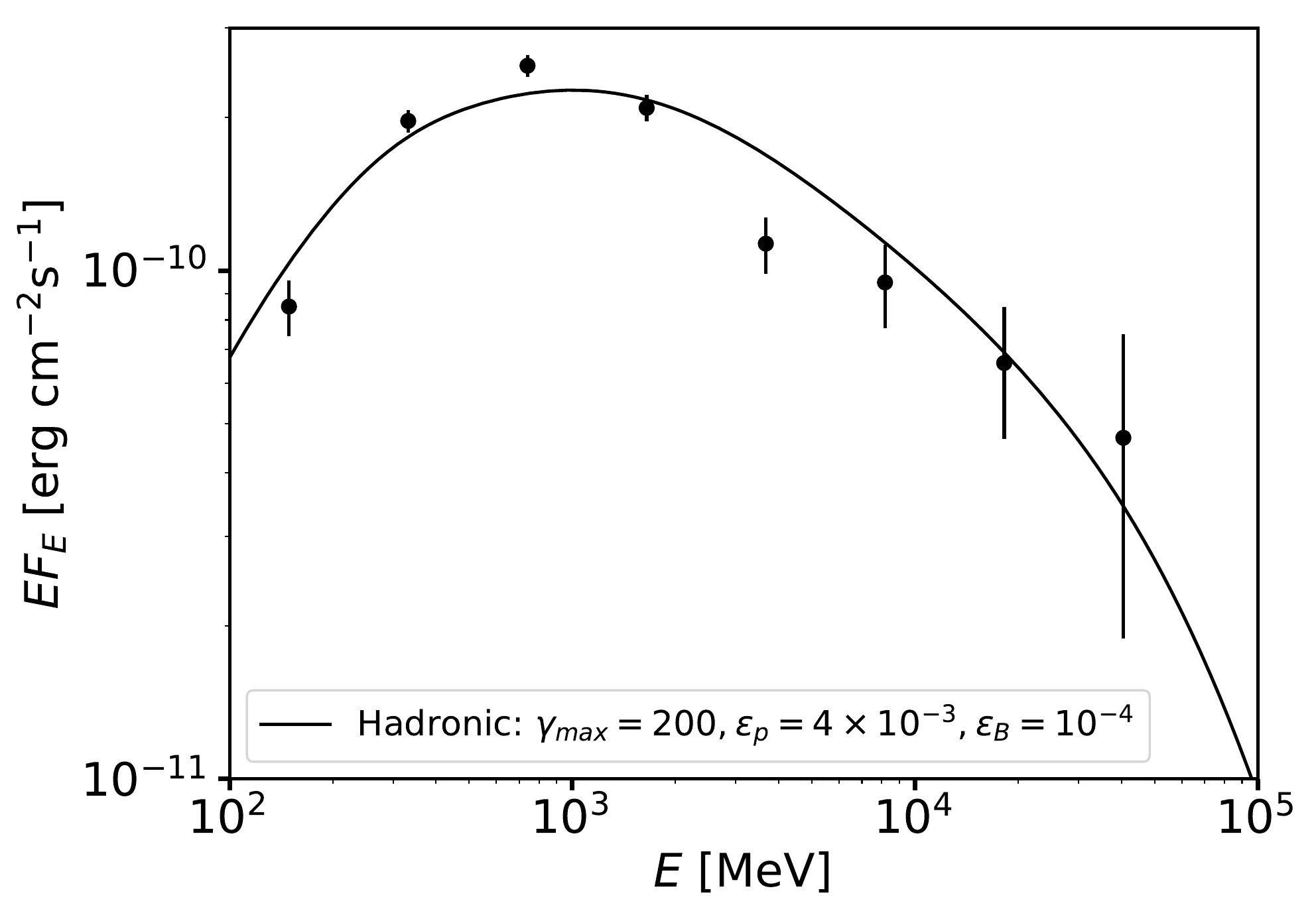}
  \includegraphics[width=0.49\textwidth]{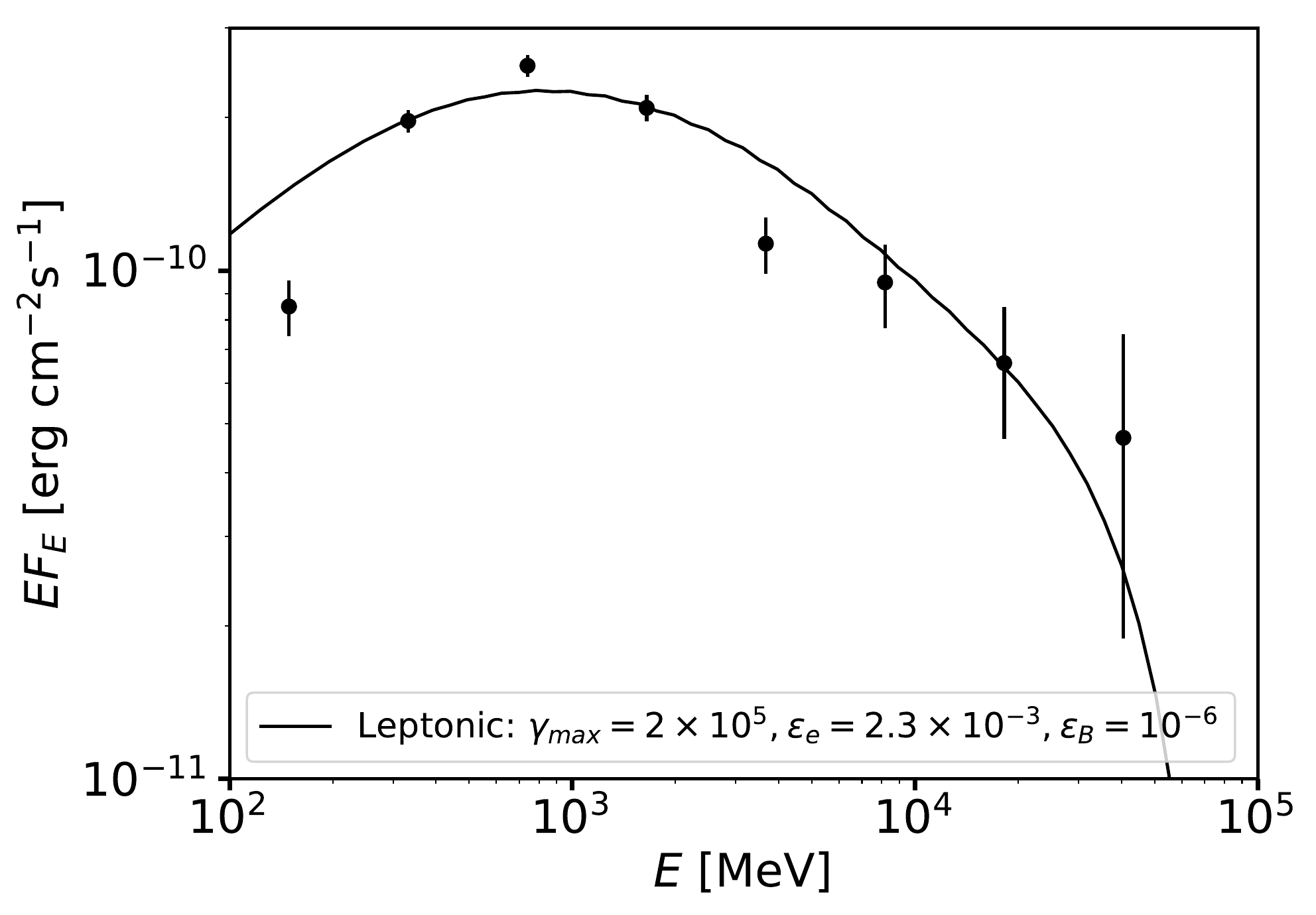}
\caption{\textbf{The hadronic (left) and leptonic (right) models overplotted on the \textit{Fermi}/LAT $\gamma$-ray time-integrated spectral energy distribution of V906~Car.} {\it Left:} the hadronic model assuming that protons are injected at the shock front with a spectrum $dN_{\rm p}/dp \propto p^{-q}$, where $p = \beta\gamma$ is the proton momentum and $q = 2.4$. {\it Right:} the leptonic model takes an electron injection spectrum of $dN_{\rm e}/d\gamma \propto \gamma^{-q}$, where $q = 1.8$. The total shock luminosity is assumed to be equal to the average optical luminosity, and we use $v_{\rm sh} = 2000$~km~s$^{-1}$ for the shock velocity. The reported errors are 1$\sigma$ uncertainties.}
\label{Fig:Hard_Lept}
\end{center}
\end{figure*}

We model $\gamma$-ray emission from shocks with a numerical code developed by Ref.\cite{Vurm_Metzger_2018}. Given the accelerated particle (electron or proton) distribution, the code follows the cooling and radiation of the relativistic particles as they are advected towards the downstream, accounting for secondary effects such as additional compression and magnetic field enhancement due to the loss of thermal pressure (see Ref.\cite{Vurm_Metzger_2018} for more details).

The results from hadronic and leptonic models for V906~Car are shown in the left and right panels of Supplementary Figure~\ref{Fig:Hard_Lept}, respectively. In the hadronic scenario, our simple 1D single-shock model with constant parameters can capture the overall spectral shape reasonably well, with the exception of the ``dip" near $3$~GeV. On the other hand, the low-energy turnover seems to be a generic feature of nova $\gamma$-ray spectra (c.f. nova V5856 Sgr\cite{Li_etal_2017_nature}) and can be more naturally reproduced by the hadronic mechanism where it is associated with the intrinsic shape of the $\pi_0$ decay spectrum. The low-energy break is also present in leptonic models and has its origin in the Coulomb (i.e. non-radiative) losses that dominate below $\sim 1$~GeV. However, the turnover in this case is more gradual as well as parameter-dependent, and is easily contaminated by sub-GeV IC radiation from electrons cooling on the intense thermal radiation field. Hence we were unable to reproduce the observed low-energy turnover by the leptonic scenario with reasonable parameters.

The leptonic scenario is also disfavored on different grounds, which mirror some of the arguments put forth in favor of hadronic models for the nova V5856 Sgr \cite{Li_etal_2017_nature}, since the parameters of the best-fit models for both leptonic and hadronic scenarios are similar in the two novae. As in V5856 Sgr, the leptonic scenario requires low postshock magnetization, $\varepsilon_B\lesssim 10^{-6}$, to avoid an efficiency problem due to strong synchrotron cooling in the compression-enhanced magnetic field. Such field strength may be too low for accelerating particles to the sufficiently high energies on the required timescale\cite{Li_etal_2017_nature,Metzger_etal_2016}. Even without synchrotron losses, the required electron acceleration efficiency, $\varepsilon_e = 2.3\times 10^{-3}$, is higher than the values inferred from particle-in-cell simulations\cite{Park_etal_2015}.

Novae are suggested to be multi-messenger sources, given that their $\gamma$-ray emission is indeed hadronic (see, e.g., Ref.\cite{Metzger_etal_2016}). If novae $\gamma$-ray emission extends above 100\,GeV, they should be accessible to future atmosphere Cherenkov telescopes, such as the Cherenkov Telescope Array (CTA). Nevertheless, detecting high-energy neutrinos ($E\,\gtrsim$\,10\,TeV) from novae by current available facilities, such as IceCube is still not feasible. However, the planned newer generation IceCube (Gen2), could make this possible\cite{Metzger_etal_2016}.

\subsection{Alternative Scenarios.}
Alternative scenarios to explain the optical--$\gamma$-ray correlation in V906~Car are if  the varying optical luminosity is originating directly from variation in the luminosity of the binary (white dwarf or accretion disk) rather than the shocks and the particle acceleration within the shocks were very efficient ($> 10$\%). Variation in the luminosity of the binary might lead to (1) the optical flares, which would not be driven by shocks in this case, and (2) time-variable, radiation-driven outflows which would in turn produce time-variable shocks, manifesting as $\gamma$-ray flares. 

Given that the $\gamma$-rays are emitted by the shock, if the change in the binary luminosity leads to an immediate acceleration of the wind, it would still require several days for subsequent ejecta with average velocities of $\gtrsim$ 1000\,km\,s$^{-1}$ to travel an average distance of a few 10$^{13}-$10$^{14}$\,cm (the distance traveled by the initial ejecta during the first 10 days) and meet slower ejecta ahead of it. However, the radiation from the white dwarf diffuses through the ejecta and emerges in the optical bands on a timescale of the order of a few minutes only. The $\gamma$-rays would thus lag the optical, possibly by several days in this scenario. 

Therefore, the observed correlation and the short time lag measured between the two light curves can best be explained if the flares in optical and $\gamma$-rays share a common origin---that is, the flares in both bands are originating from shock interactions.

\end{document}